   \documentstyle[nato]{crckapb} 
\input epsf.sty

\def\ga{\mathrel{\raise.3ex\hbox{$>$\kern-.75em\lower1ex\hbox{$\sim$}}}}
\def\la{\mathrel{\raise.3ex\hbox{$<$\kern-.75em\lower1ex\hbox{$\sim$}}}}
\def\he#1{\hbox{${}^{#1}$He}}
\def\li#1{\hbox{${}^{#1}$Li}}
\def\be#1{\hbox{${}^{#1}$Be}}
\def\b#1#2{\hbox{${}^{#1#2}$B}}
\def\beq{\begin{equation}}
\def\eeq{\end{equation}}

\begin{opening}

\title{PRIMORDIAL BIG BANG NUCLEOSYNTHESIS}
\author{K.A. OLIVE}
\institute{School of Physics and Astronomy, Theoretical Physics
Institute \\
 University of Minnesota\\
           Minneapolis MN 55455 USA}
\end{opening}

\runningtitle{PRIMORDIAL BIG BANG NUCLEOSYNTHESIS}

\begin{document}

% The \begin{document} command comes after the \end{opening}
% command.

\vskip -3in
\rightline{UMN-TH-1735/99}
\rightline{TPI-MINN-98/30}
\rightline{astro-ph/9901231}
\rightline{January 1999}
\vskip 2.2in

{\bf Abstract.} Big Bang Nucleosynthesis is the theory of the production of
the light element isotopes of D, \he3, \he4, and \li7.
After a brief review of the essential elements of the standard Big Bang
model at a temperature of about 1 MeV, the theoretical input and predictions
of BBN are discussed.  The theory is tested by the observational 
determinations of the light element abundances and the current 
status of these observations is reviewed. Concordance of 
standard model and the related observations is found over a limited 
range of the baryon-to-photon ratio, the single true parameter of the standard
model. Implications of BBN on chemical evolution, dark matter and constraints on 
particle properties will be also discussed.

\renewcommand{\thefootnote}{\fnsymbol{footnote}}
\section{Introduction}

Big Bang Nucleosynthesis\footnotetext{Summary of Lectures given at the 
Advanced School on Cosmology and Particle Physics, Peniscola, Spain, June 1998
and at the Theoretical and Observational
Cosmology Summer School , Cargese, Corsica, France, August 1998.} (BBN) is the
theory explaining the origins of the light elements D,\he3, \he4, and \li7 and
their primordial abundances. The theoretical framework for BBN is quite
simple. It is based on Friedmann-Lemaitre-Robertson-Walker cosmology
\cite{ellis} and a network of nuclear reactions.  We can further specify that
the standard BBN model refers to homogeneous nucleosynthesis in the context of
a FLRW Universe with an electroweak standard model particle content, which for
the purposes of BBN really amounts to assuming the existence of three
nearly massless and nearly stable neutrinos.  The predictions of BBN for
the abundances of the light elements are determined by running a code
which incorporates the nuclear network in a thermal (and cooling due to
the expansion of the Universe) bath.  These predictions are then be
compared with the observational determinations of the abundances.

In contrast to the theoretical side of BBN, the status of the 
observational data has changed significantly in the last several years.
There is more data on \he4 and  \li7, and data on D and
\he3  that was simply non-existent several years ago.  For the most part,
the inferred abundances of \he4 and \li7 have remained relatively fixed,
giving us  a higher degree of confidence in the assumed primordial
abundances of these isotopes as is reflected in their observational
uncertainties. Indeed, the abundances of \he4 and \li7 alone are
sufficient to probe and  test the theory and determine the
single remaining parameter in the standard model \cite{fo}, namely, the
baryon-to-photon ratio, $\eta$. In contrast, D and \he3 are highly
dependent on models of chemical evolution (\he3 is in addition dependent
on the uncertain stellar yields of this isotope). New data from quasar
absorption systems, on what may be primordial D/H is at this time
disconcordant, different measurements give different abundances. As a
consequence of the uncertainties in D and \he3,  one can hope to use the
predictions based on \he4 and \li7 in order to construct models of
galactic chemical evolution.  These results also have  important
implications for the amount of (non)-baryonic dark matter in the galaxy
and on the number of allowed relativistic degrees of freedom at the time
of BBN, commonly parameterized as $N_\nu$. 

\subsection{Standard Model Basics}

Since one of the main inputs of the theoretical side of BBN is the
standard hot big bang model, it will be useful to review some of the key
concepts as they pertain to BBN.  The metric for the FLRW model of the
Universe is specified by two quantities, the curvature constant $k$,
and the expansion scale factor $R(t)$.  At early times, the curvature is
unimportant as can be seen from the Friedmann equation for the Hubble
parameter
\beq
	H^2  \equiv \left({\dot{R} \over R}\right)^2  = 
{1 \over 3} 8 \pi G_N \rho - { k \over R^2}  + {1 \over 3} \Lambda
\label{H}
\eeq
where $\Lambda$ is the cosmological constant.
Since the density $\rho$ scales as either $R^{-3}$ (for a matter
dominated universe) or $R^{-4}$ (for a radiation dominated universe), 
this term dominates over either the curvature or the cosmological
constant.  I will ignore both in what follows. 

The critical energy density $\rho_c$ is defined
  such that $\rho =\rho_c$  for $k = 0$
\beq
	\rho_c  = 3H^2 / 8 \pi G_N		
\eeq
In terms of the present value of the Hubble parameter this is,
\beq
	\rho_c  = 1.88 \times 10^{-29} {h_o}^2  {\rm g cm}^{-3}  		
\eeq
where
\beq
	h_o  = H_o /(100 {\rm km Mpc}^{-1}   {\rm s}^{-1}  )		
\eeq
The cosmological density parameter is then defined by
\beq
	\Omega \equiv {\rho \over \rho_c} 			
\eeq
in terms of which the Friedmann equation, Eq. (\ref{H}), 
can be rewritten as (with $\Lambda = 0$)
\beq
	(\Omega - 1)H^2  = {k \over R^2}	
\label{o-1}	
\eeq
so that $k = 0, +1, -1$ corresponds to $\Omega = 1, \Omega > 1$
 and $\Omega < 1$.  
(Very) broad observational limits on $h_o$ and $\Omega$ are
\beq
0.4 \le h_o \le 1.0 \qquad 0.1 \le \Omega \le 2
\label{range}
\eeq

The value of $\Omega$, at least on relatively small scales, seems to 
depend on scale. Indeed, the contribution to $\Omega$ from visible
matter associated with stars and hot gas is quite small, $\Omega \approx
0.003 - 0.01$.  On somewhat larger scales, that of galactic halos or
small groups of galaxies, $\Omega \approx 0.02 - 0.1$.
On galaxy cluster scales, it appears that $\Omega$ may be as large as 
0.3.  And while there is some evidence, the observations are
far from conclusive in indicating a value of $\Omega$ as large as 1.
It is however possible to obtain a bound on the product, $\Omega h^2$
from 
\beq
H_o t_o = \int_0^1 ( 1 - \Omega + \Omega/x )^{-1/2} dx
\eeq
(for $\Lambda = 0$).
For $t_o > 12$Gyr, and $\Omega \le 1$, $\Omega h^2 < 0.3$
(This is true even if $\Lambda \ne 0$.)

As indicated above,  BBN  takes place during the radiation dominated
epoch which lasts roughly to the period of
recombination (somewhat earlier when dark matter is included)
 which occurs when electrons
 and protons form neutral hydrogen  through 
$e^{ -  } +  p     \rightarrow $  H  $+   \gamma $
  at a temperature  
$T_{ R} { \sim }$  few $\times 10^{3}$ K  ${ \sim}1$ eV.  For $T < T_{R}$, 
photons are decoupled while for $T > T_{ R}$,  photons are
 in thermal equilibrium.  Today, the content
 of the microwave background consists of photons with
$T_o =  2.728  \pm .002$ K \cite{cobet}.  
The energy density of photons in the background can be calculated from
\beq
 \rho_\gamma  = \int E_\gamma dn_\gamma
\label{rhog}
\eeq
 where the density of states is given by
\beq
dn_\gamma  =   {g_\gamma \over 2 
 \pi^{ 2}}[exp(E_\gamma/T)-1]^{ -  1} q^{ 2} dq 
\eeq
 and $g_\gamma = $  2 is the number of spin polarizations for the 
photon,
$E_\gamma =  q$ is just the photon energy (momentum). 
 (I am using units such that  
$\hbar =  c  = k_{ B}   =$  1 and will do so through the remainder
 of these lectures.)  
Integrating (\ref{rhog}) gives
\beq
\rho_\gamma = {\pi^2 \over 15} T^4
\eeq
 which is the familiar blackbody result.

In general, at very early times, at very high temperatures,
 other particle degrees of freedom join the radiation background when  
$T{ \sim } m_{i}$  for each  particle type $i$, if that type is brought
 into thermal equilibrium through interactions.  In equilibrium 
the energy density of a particle type $i$ is given by
\beq
 \rho_{i}  = \int E_{i} dn_{q_{i}} 
\eeq
 and
\beq 
 dn_{q_{i}} = {g_{i} \over 2  \pi^{ 2}}[exp[(E_{q_{i}} - 
\mu_{i})/T] \pm 1]^{ -1 }q^{2}dq
\eeq
where again $g_{i}$ counts the total number of degrees of freedom for type i,
\beq
 E_{q_{i}} =  \left(m_{i}^{2} + q_{i}^{ 2}\right)^{1/2}
\eeq
$\mu_{i}$ is the chemical potential if present and  $ \pm$  
corresponds to either Fermi or Bose statistics.

In the limit that  $T \gg m_{i} $  the total energy density can
 be conveniently expressed by  
\beq
 \rho {} = \left( \sum_B g_{B} + {7 \over 8} \sum_F  g_{F} \right)
   {\pi^{ 2} \over 30}  T^{4}     \equiv    {\pi^{ 2} \over 30} N(T) T^{4} 
\label{NT}
\eeq
 where $g_{B(F)} $  are the total number of boson (fermion) 
degrees of freedom and the sum runs over all boson (fermion) states with 
$m \ll T$.  The factor of 7/8 is due to the difference between
 the Fermi and Bose integrals.  Equation (\ref{NT}) defines N(T)
 by taking into account  new particle degrees 
of freedom as the temperature is raised.  

In the radiation dominated epoch,
we can obtain a relationship between
 the age of the Universe and its temperature
\beq
 t = \left({90 \over 32 \pi^3 G_{ N} N(T)}\right)^{ 1/2}  T^{ -  2} 
\label{tt1}
\eeq
 Put into a more convenient form
 \beq
 tT_{ MeV}^{ 2}  =
2.4 [N(T)]^{ -  1/2}  
\label{tt2}   
\eeq
 where t is measured in seconds and
$T_{ MeV} $  in units of MeV.

 The value of $N(T)$ at any given temperature depends
 on the particle physics model.  In the standard $SU(3) \times
SU(2)\times U(1)$  model, we can specify $N(T)$ up to 
temperatures of 0(100) GeV.
  The change in N can be seen in the following table.

\begin{table}
\caption{Effective numbers of degrees of freedom in the standard model.}
\vspace{2pt}
\begin{center}
\begin{tabular}{llc}
\hline
Temperature & New Particles \qquad
&$4N(T)$ \\
\hline\rule{0pt}{12pt}
$T < m_{ e}$   &     $\gamma$'s +   $\nu$'s & 29 \\
$m_{ e} <   T  < m_\mu$ &    $e^{\pm}$ & 43 \\
$m_\mu <  T  < m_\pi$  &   $\mu {}^{\pm}$ & 57 \\
$m_\pi <  T < T{ c}^{*}$  & $\pi$'s & 69 \\
$T_{ c} <  T  < m_{\rm strange~~~~~~~}$ \qquad &
  -  $\pi$'s + $  u,{\bar u},d,{\bar d}$ + gluons &  205 \\
$m_{ s} <  T < m_{ charm}$ & $s,{\bar s}$ & 247 \\
$m_{ c} <  T < m_\tau$ &  $c,{\bar c}$ & 289 \\
$m_\tau < T < m_{ bottom}$ & $\tau {}^{\pm}$ & 303 \\
$m_{ b} < T < m_{ W,Z}$ & $b,{\bar b}$ & 345 \\
$m_{ W,Z} <  T < m_{ top}$ & $W^{\pm}, Z$ & 381 \\
$ m_t < T < m_{Higgs}$ & $t,{\bar t}$ & 423 \\
$M_H < T $ & $H^o$ & 427 \\
\hline
\end{tabular}
\end{center}
*$T_{ c}$ 
corresponds to the confinement-deconfinement transition between
 quarks and hadrons.
 $N(T)$  is shown in Figure \ref{mark} for $T_{ c}  =  150$ and $400$ MeV.
It has been assumed that $m_{Higgs} > m_{top}$.
\end{table}

 At higher temperatures ($T \gg 100$ GeV), $N(T)$ will be model dependent.
  For example, in the minimal $SU(5)$ model, one needs to add
 to $N(T)$, 24 states for the X and Y gauge bosons, 
another 24 from the adjoint Higgs, and another 6 (in addition
to the 4 already counted in $W^\pm, Z$ and $H$) from the $ {\bar 5}$
of Higgs.
  Hence for   
$T > M_{ X}  $  in minimal $SU(5)$, $N(T)  =  160.75$. 
 In a supersymmetric model this would at least double, 
with some changes possibly necessary in the table if the
 lightest supersymmetric particle has a mass below  
$M_{ H}.$

\begin{figure}
\hspace{0.5truecm}
\epsfysize=7truecm\epsfbox{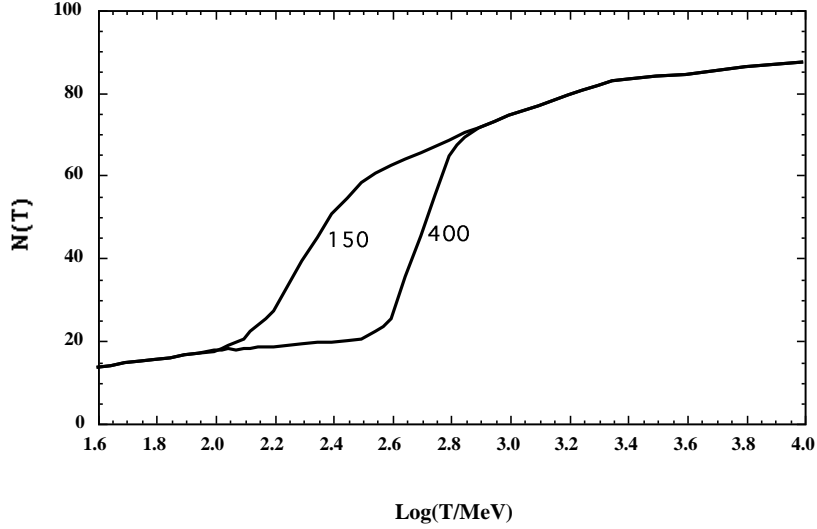}
%\vspace{-11.0truecm}
\caption{The effective numbers of relativistic degrees of freedom
as a function of temperature.}
\label{mark}
\end{figure}

The presence of a particle species in the thermal background 
assumes thermal equilibrium and hence interaction rates which
are sufficiently fast compared with the expansion rate of the 
Universe.
Roughly, this translates to the condition 
 for each particle type $i$, that some rate  
$  \Gamma {}_{ i} $  involving that type be larger than the expansion
 rate of the Universe or 
\beq
  \Gamma {}_{ i}  > H
\eeq
  in order to be in thermal equilibrium.

Examples of a  processes in equilibrium at early times 
 which drops out of equilibrium or decouples at later times
 are the processes which involve neutrinos.  
If we consider the standard neutral or charged-current interactions such as 
$e^{ +} + e^{ -  }     \leftrightarrow {}   \nu {}  +  \bar \nu $  
or $e +   \nu {}    \leftrightarrow {}  e  +   \nu {}$  etc.,
the rates for these processes can be approximated by
 \beq
 \Gamma {} =  n \langle \sigma v \rangle
 \eeq
 where  $  \langle \sigma v \rangle $  
is the thermally averaged  weak interaction cross section 
\beq
 \langle \sigma v \rangle { \sim }~0(10^{ -  2}) T^{ 2} /M_{ W}^4
\eeq
and $n$ is the number density of leptons.
 Hence the rate for these interactions is 
\beq
 \Gamma {}_{\rm wk}   { \sim  }~0(10^{ -  2}) T^{ 5}/M_{ W}^4
\label{gammaw}
\eeq 
The expansion rate, on the other hand, is just
\beq
  H  =  \left({8 \pi {}G_{ N}  \rho {} \over 3}\right)^{ 1/2}   
 =  \left({8  \pi {}^{ 3} \over 90}  N(T) \right)^{ 1/2}  T^{ 2}/M_{ P}    
 \sim 1.66 N(T)^{ 1/2}  T^{ 2}/M_{ P}.  
\label{gammae}
\eeq
The Planck mass $M_{ P} = G_N^{-1/2} =  1.22 \times 10^{19}$ GeV.

Neutrinos will be in equilibrium when  $  \Gamma {}_{\rm wk} >  H $ or
\beq
T > (500 M_{ W}^4)/M_{ P})^{ 1/3}  { \sim }~1 MeV . 
\eeq
The temperature at which these rates are equal
 is commonly referred to as the decoupling or freeze-out 
temperature and is defined by 
\beq
  \Gamma(T_{ d}) = H(T_{ d}) 
\eeq
 For temperatures $T > T_{ d}$,  
neutrinos will be in equilibrium, while for $T < T_{ d }$ 
 they will not. Basically, in terms of their interactions, 
the expansion rate is just too fast and they never  
{\em ``see"}  the rest of the matter in the Universe (or themselves).
  Their momenta will simply redshift and their effective temperature 
(the shape of their momenta distribution is not changed from that 
of a blackbody) will simply fall with  
$T { \sim  } 1/R.$

 Soon after decoupling the $e^{\pm}$  pairs in the thermal background
begin to annihilate (when $T \la m_e$).  
Because the neutrinos are decoupled, 
the energy released heats up the photon background relative 
to the neutrinos. The change in the photon temperature can be
easily computed from entropy conservation. The neutrino entropy 
must be conserved separately from the entropy of interacting particles.
 If we denote $T_>$, the temperature of photons, and $e^{\pm}$
 before annihilation, we also have $T_\nu  = T_>$  as well. 
 The entropy density of the interacting particles at $T = T_>$ is just
\beq
	s_> =  {4 \over 3}  {\rho_> \over T_>}  = 
{4 \over 3} (2 + {7 \over 2}) ( {\pi^2 \over 30} ) T^3_>	
\eeq
while at $T = T_<$, 
the temperature of the photons just after $e^{\pm}$ annihilation, 
the entropy density is
\beq
	s_< =  {4 \over 3}  {\rho_< \over T_<}  =
 {4 \over 3} (2 ) ( {\pi^2 \over 30} ) T^3_<	
\eeq
and by conservation of entropy $s_< =  s_>$  and
\beq
	(T_</T_>)^3  = 11/4 
\eeq
Thus, the photon background is at higher temperature 
than the neutrinos because the $e^{\pm}$
  annihilation energy could not be shared among the neutrinos, and
\beq
	T_\nu = (4/11)^{1/3}   T_\gamma   \simeq 1.9K 
\eeq

\subsection{Historical Perspectives}

There has always been an intimate connection between BBN and the microwave
background as a key test to the standard big bang model.
Indeed, it was the formulation of BBN which predicted the existence of 
the microwave background radiation \cite{gamo}. 
The argument is rather simple. BBN requires temperatures greater than
100 keV, which according to eqs. (\ref{tt1}) and (\ref{tt2}) corresponds
to timescales less than about 200 s. The typical cross section for the
first link in the nucleosynthetic chain is
\beq 
\sigma v (p + n \rightarrow D + \gamma) \simeq 5 \times 10^{-20} 
{\rm cm}^3/{\rm s}
\eeq
This implies that it was necessary to achieve a density
\beq
n \sim {1 \over \sigma v t} \sim 10^{17} {\rm cm}^{-3}
\eeq
The density in baryons today is known approximately from the density of
visible matter to be ${n_B}_o \sim 10^{-7}$ cm$^{-3}$ and since
we know that that the density $n$ scales as $R^{-3} \sim T^3$, 
the temperature today must be
\beq
T_o = ({n_B}_o/n)^{1/3} T_{\rm BBN} \sim 10 {\rm K}
\eeq
A pretty good estimate.  

\begin{figure}[h]
\hspace{0.5truecm}
\epsfysize=7truecm\epsfbox{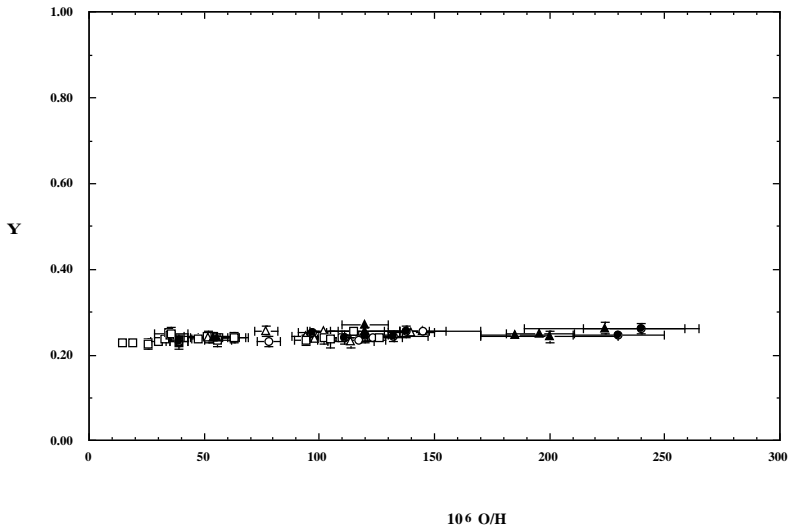}
\caption { The helium (Y) vs oxygen (O/H) abundances 
in extragalactic HII regions emphasizing the lack of low \he4 regions.  }
\label{fig:fig4gary}
\end{figure}

Despite its simplicity, BBN was criticized early on, due to its 
shortcomings in being able to produce the observed abundances of 
{\em all} of the element isotopes due primarily to the gaps in stable
nuclei at $A = 5$ and $A = 8$.  Attention was therefore turned to stellar
nucleosynthesis \cite{burb}.  However, while the elements from helium on
up can be and are produced in stars, no other astrophysical site has ever
survived for the production of deuterium \cite{rafs}.  In addition, if one
assumes that \he4 is entirely of stellar origin, one should be able to
find places in the  Universe in which the \he4 mass fraction is
substantially below 25\%. The  \he4 data shown in Figure
\ref{fig:fig4gary}, emphasizes the fact that indeed  no such region with
low \he4 has ever been observed and that (together with the need to
produce D) leads one to conclude that BBN nucleosynthesis is a necessary
component in any cosmological model. The foundations of modern BBN
continued to be laid over time \cite{hay} establishing the notions of
equilibrium and the nuclear network to obtain the abundances of D through
\li7.

\section{Theory}

Conditions for the synthesis of the light elements were attained in the
early Universe at temperatures  $T \la $ 1 MeV.  
At somewhat higher temperatures, weak interaction rates were
in equilibrium. In particular, the processes
\begin{eqnarray}
n + e^+ & \leftrightarrow  & p + {\bar \nu_e} \nonumber \\
n + \nu_e & \leftrightarrow  & p + e^- \nonumber \\
n  & \leftrightarrow  & p + e^- + {\bar \nu_e} \nonumber 
\end{eqnarray}
fix the ratio of
number densities of neutrons to protons. At $T \gg 1$ MeV, $(n/p) \simeq
1$. The energy density (dominated by radiation) in the standard model is
\beq
\rho = {\pi^2 \over 30} ( 2 + {7 \over 2} + {7 \over 4}N_\nu) T^4
\label{rho}
\eeq
from the contributions of photons, electrons and positrons, and $N_\nu$
neutrino flavors.

 As the temperature fell
and approached the point where the weak interaction rates were no longer
fast enough to maintain equilibrium, the neutron to proton ratio was given
approximately by the Boltzmann factor,
$(n/p)
\simeq e^{-\Delta m/T}$, where $\Delta m$ is the neutron-proton mass
difference. The final abundance of \he4 is very sensitive to the $(n/p)$
ratio. As in the case of the neutrino interactions discussed above,
freeze out occurs at about an MeV (slightly less than an MeV in this
case).

The nucleosynthesis chain begins with the formation of deuterium
through the process, $p+n \rightarrow$ D $+ \gamma$.
However, because the large number of photons relative to nucleons,
$\eta^{-1} = n_\gamma/n_B \sim 10^{10}$, deuterium production is delayed
past the point where the temperature has fallen below the deuterium
binding energy, $E_B = 2.2$ MeV (the average photon energy in a blackbody
is ${\bar E}_\gamma \simeq 2.7 T$).  The point being that there are many
photons in the exponential tail of the photon energy distribution with
energies $E > E_B$ despite the fact that the temperature or ${\bar
E}_\gamma$ are less than $E_B$. This can be seen by comparing the
qualitative expressions for the deuterium production and destruction
rates,
\begin{eqnarray}
\Gamma_p & \approx & n_B \sigma v \\ \nonumber
\Gamma_d & \approx & n_\gamma \sigma v e^{-E_B/T}
\end{eqnarray}
When the quantity $\eta^{-1}
{\rm exp}(-E_B/T)
\sim 1$ the rate for  deuterium destruction (D $+ \gamma \rightarrow p +
n$) finally falls below the deuterium production rate and
the nuclear chain begins at a temperature $T \sim 0.1 MeV$.

\begin{figure}[htbp]
\hspace{0.5truecm}
\epsfysize=6.0truein\epsfbox{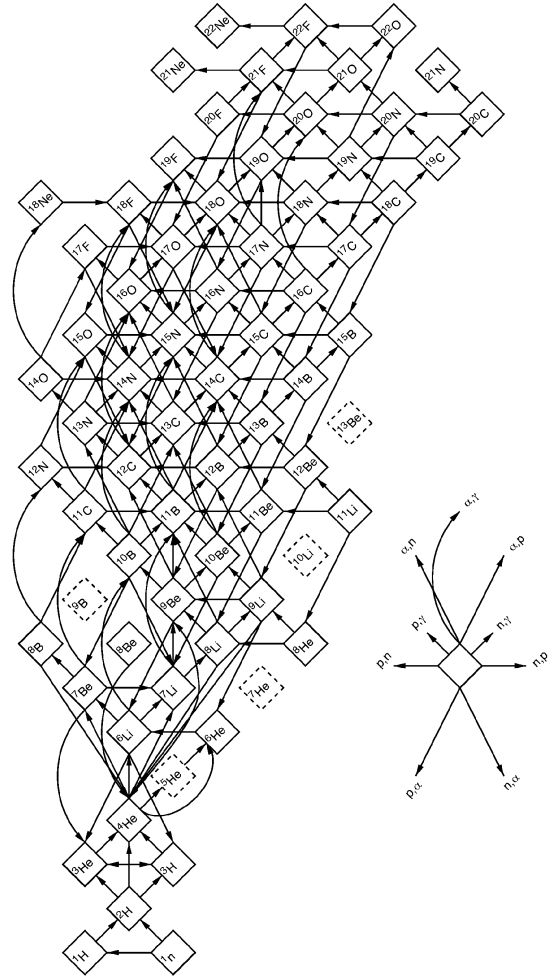}
\caption{{The nuclear network used in BBN calculations.}}
\label{net}
\end{figure}

In addition to the $p~(n, \gamma)$ D reaction, the other
major reactions leading to the production of the light
elements are:
\begin{center}
\begin{itemize}
\def\labelitemi{}
\item {D (D,~$p$) T  \qquad D ($n,\gamma$) T \qquad \he3
($n,~p$) T}
\item {D (D,~$n$) \he3 \qquad D ($p, \gamma$) \he3}
\end{itemize}
\end{center} 
Followed by the reactions producing \he4:
\begin{center}
\begin{itemize}
\def\labelitemi{}
\item {D (D, $\gamma$) \he4  \qquad  \he3 (\he3,~$2p$) \he4}
\item {D (\he3,~$p$) \he4 \qquad  T ($p,\gamma$) \he4}
\item {T (D,~$n$) \he4 \qquad  \he3 ($n,\gamma$) \he4}
\end{itemize}
\end{center} 
The gap at $A=5$ is overcome and the production of \li7
proceeds through:
\begin{center}
\begin{itemize}
\def\labelitemi{}
\item {{\ } \he3 (\he4,$\gamma$) $^7$Be } 
\item {{~~~~~~~~~~~~~~~~~~~~~} $\rightarrow
\li7 + e^+ +
\nu_e$}
\item {{\ } T (\he4,$\gamma$) \li7}
\end{itemize}
\end{center} 
The gap at $A=8$ prevents the production of other isotopes in any
significant quantity. The nuclear chain in
BBN calculations was extended
\cite{tsof} and is shown in Figure \ref{net}.

The dominant product of big bang nucleosynthesis is \he4 resulting in an
abundance of close to 25\% by mass. This quantity is easily estimated by
counting the number of neutrons present when nucleosynthesis begins.
When the weak interaction rates responsible for $n-p$
equilibrium freeze-out, at $T \approx 0.8$ MeV,
the neutron to proton ratio is about 1/6. When free neutron decays 
 prior to deuterium formation are taken into account, the ratio drops to
$(n/p) \approx 1/7$. Then simple counting yields a \he4  mass fraction
\beq
Y_p = {2(n/p) \over \left[ 1 + (n/p) \right]} \approx 0.25
\label{ynp}
\eeq

\begin{figure}[thbp]
\hspace{0.5truecm}
\vskip 2cm
\epsfysize=5.0truein\epsfbox{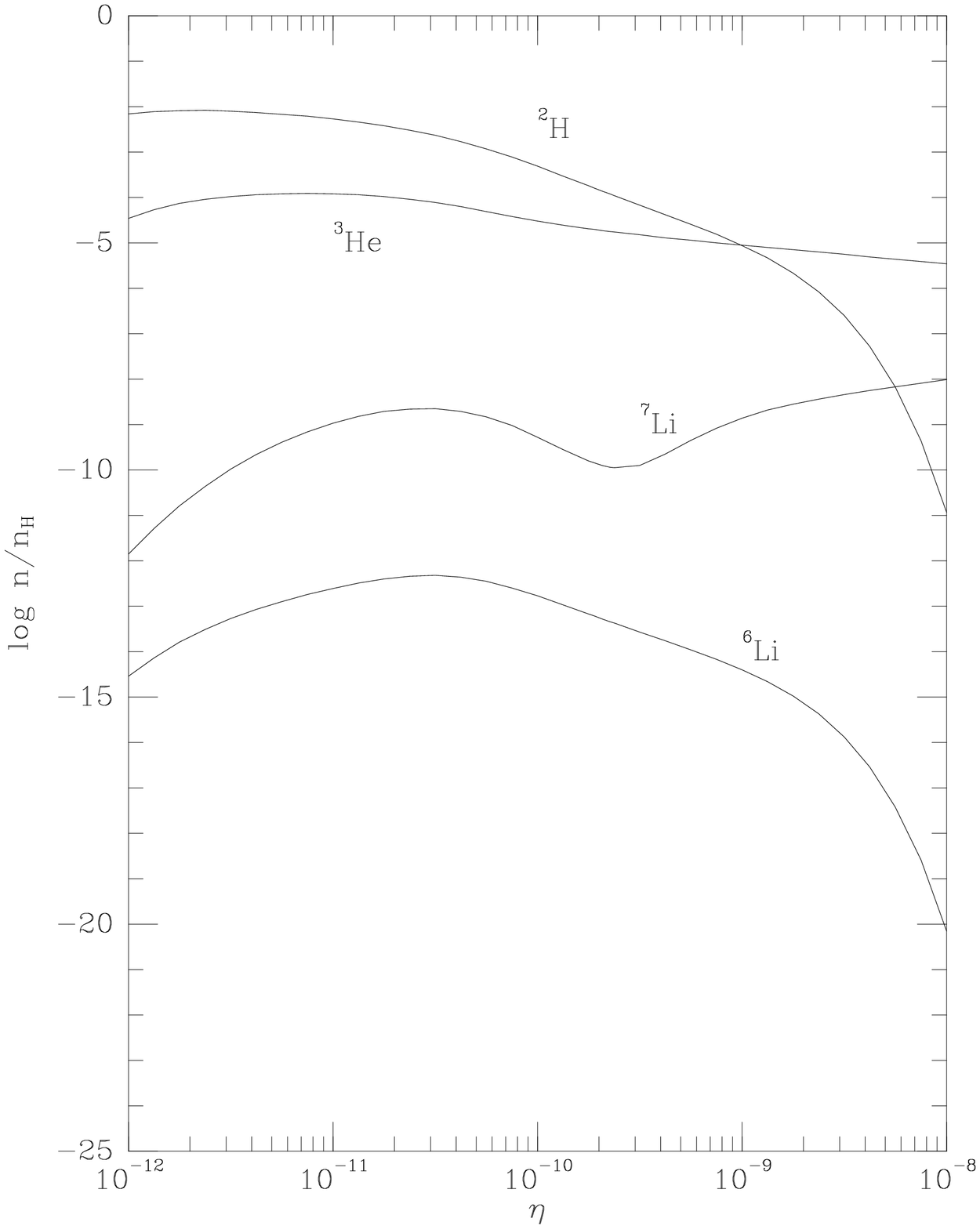}
\caption{{The light element abundances from big bang
nucleosynthesis as a function of $\eta$, including \li6.}}
\label{fig:tsof1}
\end{figure}
\begin{figure}[htbp]
\hspace{0.5truecm}
\vskip 2.0cm
\epsfysize=5.0truein\epsfbox{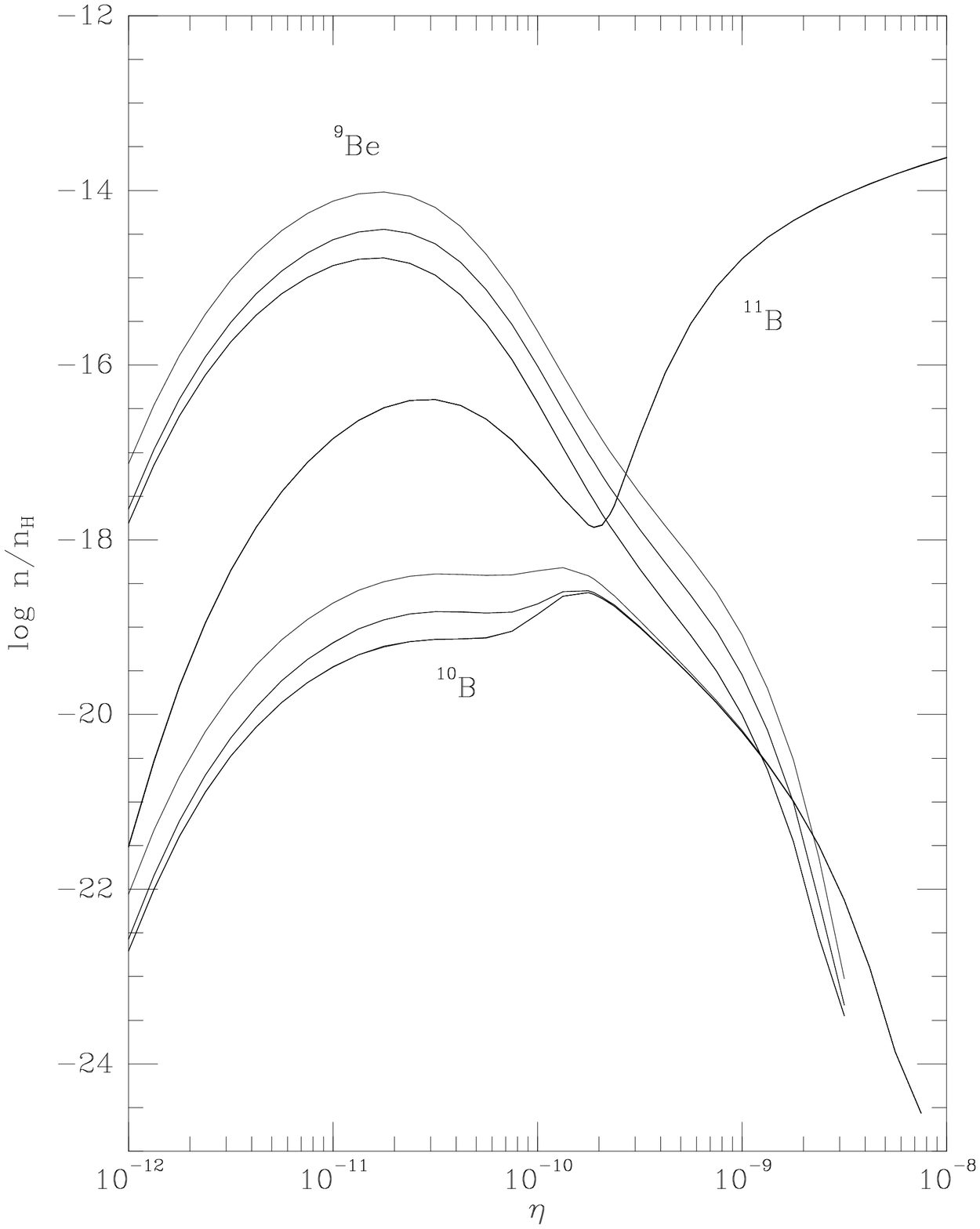}
\caption{{The intermediate mass element abundances from big bang
nucleosynthesis as a function of $\eta$.}}
\label{fig:tsof2}
\end{figure}

In the standard model with $N_\nu = 3$, there is basically one free
parameter in BBN, namely the baryon to photon ratio, $\eta$. 
As we have seen above, the value of $\eta$ controls the onset of
nucleosynthesis through the deuterium bottleneck.
For larger values of $\eta$, the quantity 
$\eta^{-1} {\rm exp}(-E_B/T)$ is smaller, and hence the nuclear chain
may begin at a higher temperature.   Remember also that a key ingredient in determining the final
mass fraction of \he4, is $(n/p)$ [see eq. (\ref{ynp})] and that the final
value of $(n/p)$ was determined by the time at which nucleosynthesis
begins, thus controlling the time available for free decays after freeze
out.  If nucleosynthesis begins earlier, this leaves less time for
neutrons to decay and the value of $(n/p)$ and hence $Y_p$ is increased.
But because the $(n/p)$ ratio is only
weakly dependent on $\eta$, the \he4 mass fraction is relatively
flat as a function of $\eta$. When we go beyond the standard model, the
\he4 abundance is very sensitive to changes in the expansion rate which 
can be related to the effective number of neutrino flavors as will
be discussed below. Lesser amounts of the other light elements are produced:
D and \he3 at the level of about $10^{-5}$ by number, 
and \li7 at the level of $10^{-10}$ by number. These abundances (along with
\li6) are shown in Figure \ref{fig:tsof1} \cite{tsof}.  In Figure
\ref{fig:tsof2},  the produced abundances of the intermediate mass
isotopes \be9, \b10,
\b11 are also shown.  These abundances are far below the observed values
and it is believed that these isotopes are formed in cosmic
ray nucleosynthesis.

It is perhaps convenient at this time to note that the value of $\eta$ is
directly related to the fraction of $\Omega$ in baryons.  Indeed, one can
write
\beq
\Omega_B h^2 = 3.67 \times 10^7 \eta (T_0/2.728 {\rm K})^3
\eeq
where $T_0$ is the present temperature of the microwave background.

Historically, it has been common to refer to two other parameter in BBN,
the neutron mean life and the number of neutrino flavors. The neutron
mean life is now very well determined and its remaining uncertainty can
be treated simply as an uncertainty in the calculated abundance of \he4.
Although the number of neutrino flavors has also been fixed
experimentally, BBN is sensitive to the number of light degrees of
freedom whether or not they interact weakly.  It is often convenient to
refer to these degrees of freedom as neutrino equivalents.  By increasing
$N_\nu$ in eq. (\ref{rho}), one increases the expansion rate $H \propto
\sqrt{\rho}$. As a result, the weak interactions freeze-out at a higher
temperature (see eqs. (\ref{gammaw}) and (\ref{gammae})).  This leads
once again to a higher value for $(n/p)$ and a higher \he4 mass fraction.
This effect will be treated in more detail in the last lecture.

\begin{figure}[htbp]
\hspace{0.5truecm}
\epsfysize=5.5truein\epsfbox{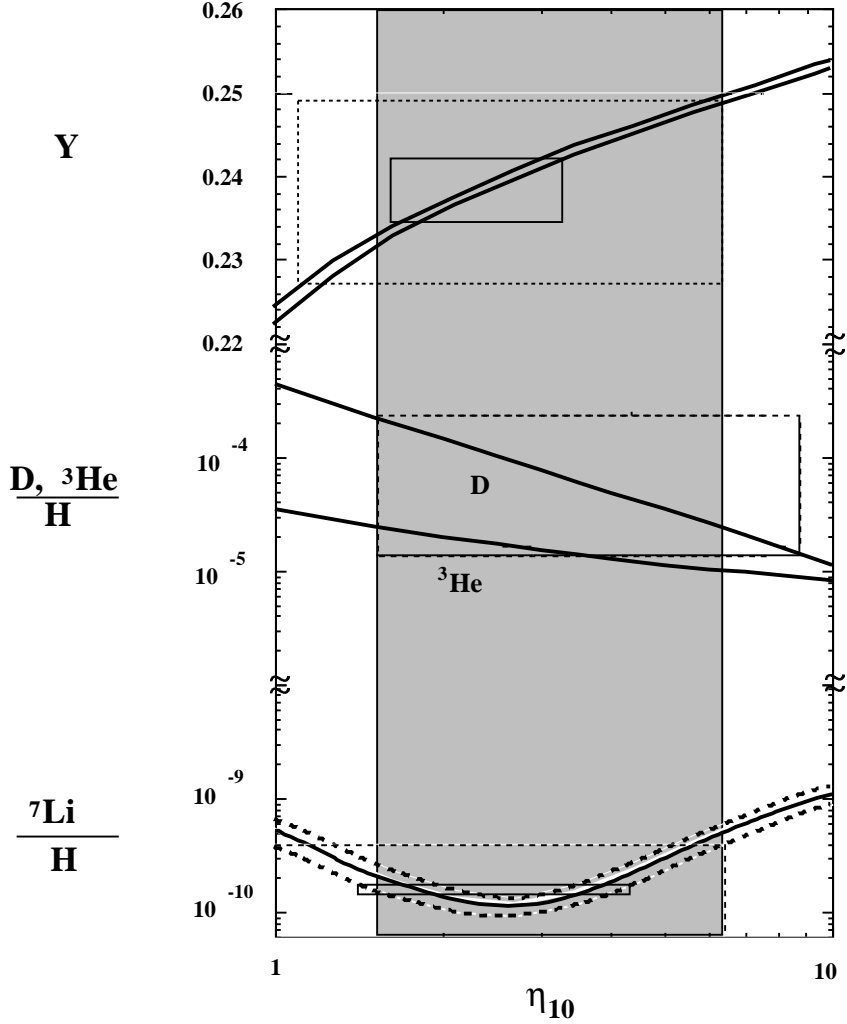}
\caption{{The light element abundances from big bang
nucleosynthesis as a function of $\eta_{10} = 10^{10}\eta$.}}
\label{nuc8}
\end{figure}

 For the
comparison with the observations, I will use the resulting abundances of
the light elements shown in Figure \ref{nuc8}, which concentrate on the
range in
$\eta_{10}$ between 1 and 10.  The curves for the \he4 mass fraction,
$Y$, bracket the computed range based on the uncertainty of the neutron
mean-life which  has been taken as \cite{rpp} $\tau_n = 887 \pm 2$ s. 
 Uncertainties in the produced \li7 
abundances have been adopted from the results in Hata et al.
\cite{hata1}. Uncertainties in D and
\he3 production are small on the scale of this figure. 
The  boxes correspond
to the observed abundances and will be discussed below. 

\section{Abundances}

\subsection{\he4}

{\ }\he4 is produced in stars, therefore to be able to extract a
primordial abundance of \he4, it is advantageous to make abundance
measurements in systems of very low metallicity. Low metallicity or low
abundances of C, N, and O relative to the solar abundances of these
elements would indicate that the level of stellar processing is minimized.
The \he4 abundance in very low metallicity regions is best determined from
observations  of HeII
$\rightarrow$ HeI recombination lines in extragalactic HII 
(ionized hydrogen) regions.
There is now a good collection of abundance information on the \he4 mass
fraction, $Y$, O/H, and N/H in over 70 
\cite{p,evan,iz} such regions. 
In an extensive study based on the data in \cite{p,evan}, it was found
\cite{osa} that the data is well represented by a linear correlation
for Y vs. O/H and Y vs. N/H.  It is then expected that the primordial
abundance of \he4 can be determined from the intercept of that
relation.   A detailed analysis of the data including that in \cite{iz}
found an intercept corresponding to a primordial abundance  
 $Y_p  = 0.234 \pm 0.002 \pm 0.005$ \cite{ost3}.
The stability of this fit was verified by a 
statistical bootstrap analysis \cite{osc} showing that the fits were not
overly sensitive to any particular HII region.

To make use of the \he4 data, it is crucial to obtain
high quality and very low metallicity data.
In principle,  any one HII region (with 
non-zero metallicity) should provide an upper limit to $Y_p$
since some stellar processing has taken place augmenting the primordial
value. Thus the determination of $Y_p$ by an extrapolation to zero 
metallicity could be avoided by the observations of either 
low metallicity or low helium HII regions.  
For a very low metallicity HII region such an upper limit may even 
provide a reasonable estimate of $Y_p$. 

\begin{figure}
\vskip -1cm
%\begin{center}
\hspace{3truecm}
\epsfysize=11truecm\epsfbox{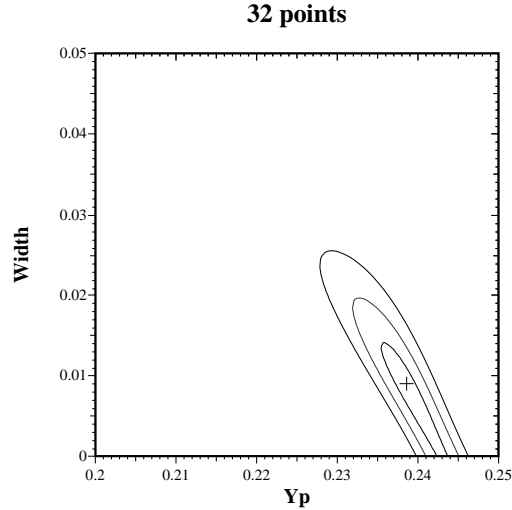}
%\end{center}
\vskip -1.0cm

\caption { \baselineskip=2ex Equal likelihood 1, 2, and 3 $\sigma$
contours in the $w - Y_p$ plane.  The cross shows the position of the
peak of the likelihood function.  }
\label{fig:hos}
\end{figure}

Another way to avoid an extrapolation to zero metallicity (though such an
extrapolation is in fact quite minimal given the low metallicity data
available), one can perform a Bayesian analysis \cite{hos} in which one
makes no other assumption other than the observed \he4 abundance is
greater than or equal to the primordial abundance. That is, one can
consider three quantities: $Y_T$, the true \he4 abundance in an H II
region about which the observed abundance $Y_O$ is distributed. 
Both of these differ from the primordial abundance $Y_p$ and the
only prior assumed is that for each object, $Y_T \ge Y_p$. If we
assume that the true abundance differs from the primordial
abundance by no more than $w$, we can derive a total likelihood function
by integrating out the unknown $Y_T$.  We can then plot the equal
likelihood contours as a function of $Y_p$ and $w$.  This is shown in
Figure \ref{fig:hos},
where the 32 points of lowest metallicity have been used to calculate the
likelihood function. The peak occurs at $Y_p = 0.238$ and the most likely
width is $w = 0.009$. The 95\% CL upper limit to $Y_p$ in this case is
0.245.  For further details on this approach see \cite{hos}. The
data used in Figure \ref{fig:hos} represents an update of that
work and includes the data of ref. \cite{iz2}.

\begin{figure}
\hspace{0.5truecm}
\epsfysize=7truecm\epsfbox{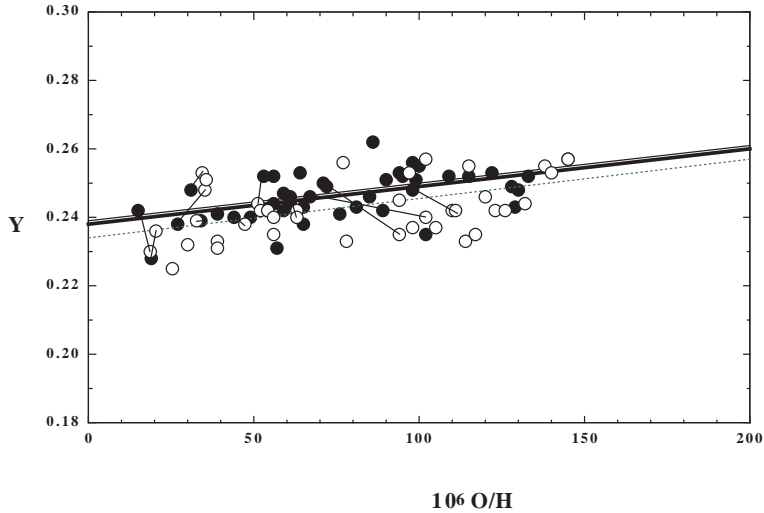}
%\vspace{-11.0truecm}

\caption { \baselineskip=2ex The helium (Y) and oxygen (O/H) abundances 
 in extragalactic HII regions, 
from refs. \protect\cite{p,evan} (open circles), and from ref.
\protect\cite{iz2} (filled circles).   Lines connect the  same regions
observed by different groups.  }
\label{fig:figost36}
\end{figure}

Although the above estimates on $Y_p$ are consistent with those based on 
a linear extrapolation of the data,
it has been claimed that the new data in refs. \cite{iz} and \cite{iz2}
leads to a significantly higher value for $Y_p$ (in excess of 24\%).
The higher values of $Y_p$, quoted by Izotov and Thuan are based
on their \he4 abundances derived by their method of determining all of the
parameters from a set of 5 helium recombination lines. This gives 
$Y_p = 0.2444 \pm 0.0015 + (44 \pm 19)$O/H.  
However, as argued in \cite{ost3} there are inherent
uncertainties in this method which are not reflected in the error budget.
For this reason and because we can more easily compare their data with
previous data, we use their results which are based on S II
densities. 
These results are entirely consistent with the data in
refs. \cite{p,evan} as can be seen in Figure \ref{fig:figost36} where 
the Y versus O/H data from
refs. \cite{p,evan} (open circles) is shown along with the newer 
data (filled circles, from ref. \cite{iz2}). The fit to the open
circles is shown by the dashed line with intercept 0.234, the fit to
the filled circles is shown by the this solid line (barely visible)
with intercept 0.239.   Combining all the data one finds
\cite{fdo2}
a \he4 mass fraction based on 73 distinct HII regions
\beq
Y_p = 0.238 \pm 0.002 \pm 0.005
\label{he4}
\eeq
The first uncertainty is purely statistical and the second uncertainty is
an estimate of the systematic uncertainty in the primordial abundance
determination \cite{ost3}. The fit to all the data is shown by the thick line with intercept given by
Eq. (\ref{he4}) above.  The small errors quoted in \cite{iz2} account for
the total fit being skewed to the higher value of $Y_p$. The solid box for
\he4 in Figure \ref{nuc8} represents the range (at 2$\sigma_{\rm stat}$)
from (\ref{he4}). The dashed box extends this by including the systematic
uncertainty. A
somewhat lower primordial abundance of  $Y_p = 0.235 \pm .003$ is found
by restricting to the 36 most metal poor regions \cite{fdo2}.  

The primordial \he4 abundance can also be determined by examining the
correlation between $Y$ and N/H.  Indeed in all but one of the H II
regions, N/H data is also available.  However, unless N/H is directly
proportional to O/H, it is not clear that a linear $Y$ vs. N/H fit
should give the same result.  Indeed, the proportionality of N/H to O/H
(or in other terms the primary vs secondary nature of nitrogen)
has been studied \cite{osa,ost3,fdo2}.  Unfortunately from a theoretical
point of view this question lies in the realm of very uncertain yields
for nitrogen in AGB stars. The data indicate that N is mostly primary. 
Though the secondary contribution may be responsible for yielding
systematically higher intercepts for $Y$ vs N/H relative to $Y$ vs.
O/H, however the difference is small $\la 0.003$. 

Finally, it also of interest to
test our understanding of the slope in the $Y$ vs. O/H data. The data
overall show a relatively steep slope $\Delta Y/\Delta O \simeq 110 \pm
25$.  Models of chemical evolution typically give a much smaller value of
about 20 and even in models with outflow (material ejected from the
galaxy) the slopes only go up to about 60.  This question, like the N vs.
O question is highly sensitive to very uncertain theoretical yields
\cite{fdo2}.

\subsection{\li7}

The \li7 abundance
is also reasonably well known.
 In old,
hot, population-II stars, \li7 is found to have a very
nearly  uniform abundance \cite{sp}. For
stars with a surface temperature $T > 5500$~K
and a metallicity less than about
1/20th solar (so that effects such as stellar convection may not be important),
the  abundances show little or no dispersion beyond that which is
consistent with the errors of individual measurements.
Indeed, as detailed in ref. \cite{mol}, much of the work concerning
\li7 has to do with the presence or absence of dispersion and whether
or not there is in fact some tiny slope to a [Li] = $\log$ \li7/H + 12 vs.
T or [Li] vs. [Fe/H]  relationship ([Fe/H] is the log of the Fe/H ratio
relative to the solar value).

\begin{figure}
\hspace{0.5truecm}
\epsfysize=7truecm\epsfbox{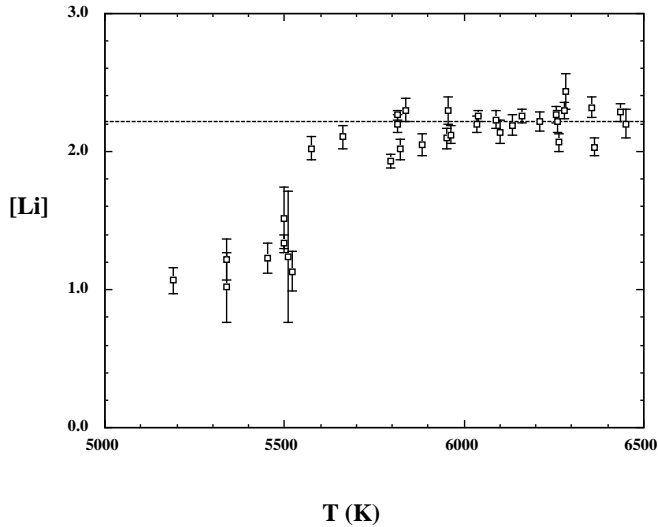}
\caption{The Li abundance in halo stars with [Fe/H] $<$ -1.3,
as a function of surface temperature. The dashed line shows the value of
the weighted mean of the plateau data.}
%\vspace{-11.0truecm}
\label{fig:lit}
\end{figure}

\begin{figure}
\hspace{0.5truecm}
\epsfysize=7truecm\epsfbox{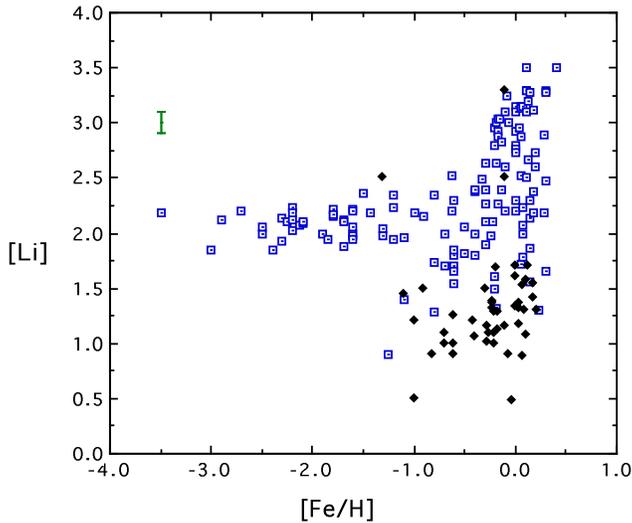}
\caption{The Li abundance in halo stars with $T > 5500$ K,
as a function of metallicity. Filled diamonds represent upper limits.}
\label{fig:life}
\end{figure}

There is \li7 data from nearly 100 halo stars, from a 
 variety of sources. When the Li data from stars with [Fe/H] $<$ 
-1.3 is plotted as a function of surface temperature, one sees a 
plateau emerging for $T > 5500$ K as shown in Figure \ref{fig:lit} for 
the data taken from ref. \cite{mol}.
As one can see from the figure, at high temperatures, where the convection
zone does not go deep below the surface, the Li abundance is uniform.
At lower temperatures, the surface abundance of Li is depleted as
Li passes through the hotter interior of the star and is destroyed.
The lack of dispersion in the plateau region is evidence that this abundance
is indeed primordial (or at least very close to it).  
Another way to see the plateau is to plot the Li abundance data as a 
function of metallicity, this time with the restriction that
$T > 5500$ K as seen in Figure \ref{fig:life}.
In this case, the plateau emerges at low [Fe/H] as would be expected.
At higher [Fe/H], the convection zone remains near the surface 
only for much hotter stars. Thus, for [Fe/H] $>$ -1.3, the effects
of depletion are seen.  Also apparent in this figure is that at higher
metallicities there is evidence for the production of Li which
rises by over an order of magnitude at solar metallicity.

I will use the value given in ref. \cite{mol} 
as the best estimate
for the mean \li7 abundance and its statistical uncertainty in halo stars 
\beq
{\rm Li/H = (1.6 \pm 0.1 ) \times 10^{-10}}
\label{li}
\eeq
The Li abundance determination is sensitive to stellar parameters
such as the assumed surface temperature, the metallicity and the surface
gravity. The greatest model dependence is on the conversion of the
observed colors (B-V) to temperature.  For example, in a sample of 55
stars taken from the papers of ref. \cite{sp}, one finds [Li] = $2.08 \pm
0.02$. From Thorburn's\cite{thor} sample of 74 stars one finds [Li] =
$2.26 \pm 0.01$. I believe that much of this systematic uncertainty has
now been relieved by newer methods for determining the stellar parameters
and the Li abundance. The two papers in ref. \cite{mol}, are based on
different methods.  The first uses temperatures which are determined
by observations of Balmer lines \cite{fuhr} and the second employs
the infra-red flux method \cite{alonso}.  The data sets in these
papers which contain 24 and 41 stars respectively, both give [Li] =
$2.21 
\pm 0.01$. The Li abundance used in (\ref{li}) corresponds to this
value.   
In fact when correlated with respect to either temperature or iron, the
data in the latter paper of \cite{mol} shows no statistical trend.  With
respect to temperature, the data give [Li] $= 1.28 \pm 0.43 + (0.015 +
0.007)T/100$K, indicating a change in [Li] well within the data over the
observed temperature range.  With respect to iron, no slope is found 
[Li] $=  2.17 \pm 0.07 + (-0.018 \pm 0.031)$[Fe/H]. The variance in the
plateau data is less than 0.01. 
The solid box for \li7 in Figure \ref{nuc8} represents
the 2$\sigma_{\rm stat}$ range from (\ref{li}).

There is however an important source of systematic error due to the
possibility that Li has been depleted in these stars from their initial
abundance. These uncertainties are however limited. 
As was noted above, the lack of dispersion in the Li data limits the
amount of depletion.  In addition,  standard stellar models\cite{del}
predict that any depletion of \li7 would be  accompanied by a very severe
depletion of \li6.  Until recently, 
\li6 had never been observed in
hot pop II stars. The observation\cite{li6o} of 
\li6 (which turns out to be
consistent with its origin in cosmic-ray nucleosynthesis 
and with a small amount
of depletion as expected from standard stellar models) 
is  another good indication that
\li7 has not been destroyed in these stars\cite{li6,pin,vcof}.

Aside from the big bang, Li is  produced together with Be and B in cosmic
ray spallation of C,N,O by protons and $\alpha$-particles.  Li is also
produced by  $\alpha-\alpha$ fusion.  Be and B have been observed in
these same pop II stars and in particular there are a dozen or so stars
in which both Be and \li7 have been
observed.  Thus Be (and B though there is still a paucity of
data) can be used as a consistency check on primordial Li \cite{fossw}. 
Based on the Be abundance found in these stars, 
one can conclude that no more than 10-20\% of 
the \li7 is due to cosmic ray nucleosynthesis leaving the remainder
(an abundance near $10^{-10}$) as primordial.
The dashed box in Figure \ref{nuc8}, accounts for
 the possibility that as much as half
of the primordial \li7 has been
destroyed in stars, and that as much as 20\% of the observed \li7   may
have been produced in cosmic ray collisions rather than in the Big Bang.
 For \li7, the uncertainties are clearly dominated by
systematic effects.

\subsection{D}

Turning to D/H, we have three basic types of abundance information:
1) ISM data, 2) solar system information, and perhaps 3) a primordial
abundance from quasar absorption systems.  The best measurement for ISM D/H
is \cite{linetal}
\beq
{\rm (D/H)_{ISM}} = 1.60\pm0.09{}^{+0.05}_{-0.10} \times 10^{-5}
\eeq
Because there are no known astrophysical sites for the production of
deuterium, all observed D must be primordial. As a result,
a firm lower bound from deuterium establishes an upper bound on $\eta$
which is robust and is shown by the lower right of the solid box
in Figure \ref{nuc8}.
 The solar abundance of D/H is inferred from two
distinct measurements of \he3. The solar wind measurements of \he3 as well as 
the low temperature components of step-wise heating measurements of \he3 in
meteorites yield the presolar (D + \he3)/H ratio, as D was 
efficiently burned to
\he3 in the Sun's pre-main-sequence phase.  These measurements 
indicate that \cite{scostv,geiss}
\beq
{\rm \left({D +~^3He \over H} \right)_\odot = (4.1 \pm 0.6 \pm 1.4) \times
10^{-5}}
\eeq
 The high temperature components in meteorites are believed to yield the true
solar \he3/H ratio of \cite{scostv,geiss}
\beq
{\rm \left({~^3He \over H} \right)_\odot = (1.5 \pm 0.2 \pm 0.3) \times
10^{-5}}
\label{he3}
\eeq
The difference between these two abundances reveals the presolar D/H ratio,
giving,
\beq
{\rm (D/H)_{\odot}} \approx (2.6 \pm 0.6 \pm 1.4) \times 10^{-5}
\eeq

It should be noted that measurements of surface abundances 
of HD on Jupiter
show a somewhat higher value for D/H,  D/H = $5 \pm 2 \times 10^{-5}$ 
\cite{nie}. If this value is confirmed and if fractionation
does not significantly alter the D/H ratio (as it was suspected to for 
previous measurements involving CH$_3$D), it may have an important 
impact on galactic chemical evolution models.  This value
is marginally consistent with the inferred meteoritic values. 

Finally, there have been several reported measurements of 
D/H in high redshift quasar absorption systems. Such measurements are in
principle capable of determining the primordial value for D/H and hence
$\eta$, because of the strong and monotonic dependence of D/H on $\eta$.
However, at present, detections of D/H  using quasar absorption systems
do not yield a conclusive value for D/H.  As such, it should be cautioned 
that these values may not
turn  out to represent the true primordial value and it is very unlikely 
that both are primordial and indicate an inhomogeneity \cite{cos2}
(a large scale inhomogeneity of the magnitude required to placate all
observations is excluded by the isotropy of the microwave background
radiation). The first of these measurements
\cite{quas1} indicated a rather high D/H ratio, D/H $\approx$ 1.9 -- 2.5
$\times 10^{-4}$.  Other  high D/H ratios were reported in \cite{quas3}. 
More  recently, a similarly high value of D/H = 2.0 $\pm 0.5 \times
10^{-4}$ was reported in a relatively low redshift system (making it less
suspect to interloper problems) \cite{webb}. However, there are reported
low values of D/H in other such systems
\cite{quas2} with values of D/H originally reported as low as
$\simeq 2.5
\times 10^{-5}$, significantly lower than the ones quoted above. 
The abundance in these systems has been revised upwards to about 3.4 $\pm
0.3 \times 10^{-5}$ \cite{bty}. I will not enter into the debate as to
which if any of these observations may be a better representation of the
true primordial D/H ratio.  I only note that it remains a highly
contested issue \cite{bty,swc}
The range of quasar absorber D/H is shown by the dashed box in Figure
\ref{nuc8}.

There are also several types of \he3 measurements. As noted above, meteoritic
extractions yield a presolar value for \he3/H as given in Eq. (\ref{he3}).
In addition, there are several ISM measurements of \he3 in galactic HII
regions \cite{bbbrw} which show a wide dispersion which may be indicative 
of pollution or a bias \cite{orstv}
\beq
 {\rm \left({~^3He \over H} \right)_{HII}} \simeq 1 - 5 \times 10^{-5}
\eeq
There is also a recent ISM measurement of \he3 \cite{gg}
with
\beq
 {\rm \left({~^3He \over H} \right)_{ISM}} = 2.1^{+.9}_{-.8} \times 10^{-5}
\eeq
  Finally there are observations of \he3 in planetary
nebulae \cite{rood} which show a very high \he3 abundance of 
\he3/H $\sim 10^{-3}$.

Each of the light element isotopes can be made consistent with theory for a
specific range in $\eta$. Overall consistency of course requires that
the range in $\eta$ agree among all four light elements.
However, as will be argued below D and \he3 are far more sensitive to 
chemical evolution than \he4 or \li7 and as such the direct comparison
between the theoretical predictions of the primordial abundances of
D and \he3 with the observational determination of their abundances is far more 
difficult.  Therefore in what follows I will for the most part 
restrict the comparison between
theory and observation to the two isotopes who suffer the least from the
effects of chemical evolution.

\section{Chemical Evolution}

Because we can not directly measure the primordial abundances of any of the
light element isotopes, we are required to make some assumptions concerning
the evolution of these isotopes. As has been discussed above, 
\he4 is produced in stars along with oxygen and nitrogen.
\li7 can be destroyed in stars and produced in several
(though still uncertain) environments. D is totally destroyed in the star 
formation process and \he3 is both produced and destroyed in stars with
fairly uncertain yields. It is therefore preferable, if possible
to observe the light element isotopes in a low metallicity 
environment. Such is the case with \he4 and \li7, and we can be fairly
assured that the abundance determinations of these isotopes are close to 
primordial.  If the quasar absorption system measurements of D/H stabilize,
then this too may be very close to a primordial measurement.  Otherwise,
to match the solar and present abundances of D and \he3 to their 
primordial values requires a model of galactic chemical evolution.

 The main inputs to chemical evolution models are:
 1) The initial mass function, $\phi(m)$, indicating the 
distribution of stellar masses. Typically, a simple
power law form for the IMF is chosen, $\phi(m) \sim m^{-x}$,
with $x \simeq -2.7$.  This is a fairly good representation of the
observed distribution, particularly at larger masses.
 2) The star formation rate, $\psi$. Typical choices for a SFR
are $\psi(t) \propto \sigma$ or $\sigma^2$ or even a straight exponential
$e^{-t/\tau}$.  $\sigma$ is the fraction of mass in gas, 
$M_{\rm gas}/M_{\rm tot}$. 3) The presence
of infalling or outflowing gas; and of course 4) the stellar yields.  It is 
the latter, particularly in the case of \he3, that is the cause for
so much uncertainty. Chemical evolution models simply set up a series of 
evolution equations which trace desired quantities.  For example,
the mass in gas and the SFR evolve through a relation such as 
\beq
{dM_{\rm gas} \over dt} = -\psi(t) + e(t) + i(t) - o(t)
\eeq
where $e$ represents the amount of gas ejected from stars, $i$
is the gas infall rate, and $o$ is the gas outflow rate.
The ejection rate is in turn given by
\beq
e(t) = \int (m-m_R) \phi(m) \psi(t-\tau(m)) dm
\label{eject}
\eeq
where $m_R$ is the remnant mass (a function of the stellar mass $m$ as
well)  and $\tau(m)$ is the stellar lifetime. If we ignore
$\tau(m)$, then the ejection rate is simply proportional to the
star formation rate $\psi$, $e(t) = R \psi$.  $R$ is referred to as
the return fraction and this approximation is known as the
instantaneous recycling approximation (IRA).  Similar equations can be
developed which trace the abundances of the element isotopes
\cite{bt}. Neglecting both infall and outflow,
these take the form
\beq
{d(X M_{\rm gas}) \over dt} = -\psi(t) X + e_X(t)
\label{devol}
\eeq
where $X$ is the mass fraction of a particular element of interest and
$e_X$ is the mass fraction of the element ejected in the death of a star.
In the case of deuterium, $e_D = 0$.

As one can see from (\ref{devol}) deuterium is always a monotonically
decreasing function of time in chemical evolution models.  The degree to
which  D is destroyed, is however a model dependent
question which depends sensitively on the IMF and SFR.
The evolution of \he3 is however considerably more complicated.
Stellar models predict that substantial amounts of \he3 are
produced in stars between 1 and 3 M$_\odot$. For M $<$ 8M$_\odot$, Iben and
Truran \cite{it} calculate
\beq
(^3{\rm He/H})_f = 1.8 \times 10^{-4}\left({M_\odot \over M}\right)^2 
+ 0.7\left[({\rm D+~^3He)/H}\right]_i
\label{it}
\eeq
so that for example, when $\eta_{10} = 3$, 
((D + \he3)/H)$_i = 9 \times 10^{-5}$, and the ratio of the final abundance
of \he3/H to the initial (D + \he3)/H abundance denoted by $g_3$ is 
$g_3(1 $M$_\odot$) = 2.7. The \he3 abundance is nearly tripled.
It should be emphasized that this prediction is in
fact consistent with the observation of high \he3/H in planetary nebulae
\cite{rood}.

\begin{figure}
\hspace{0.5truecm}
\epsfysize=8truecm
\epsfbox{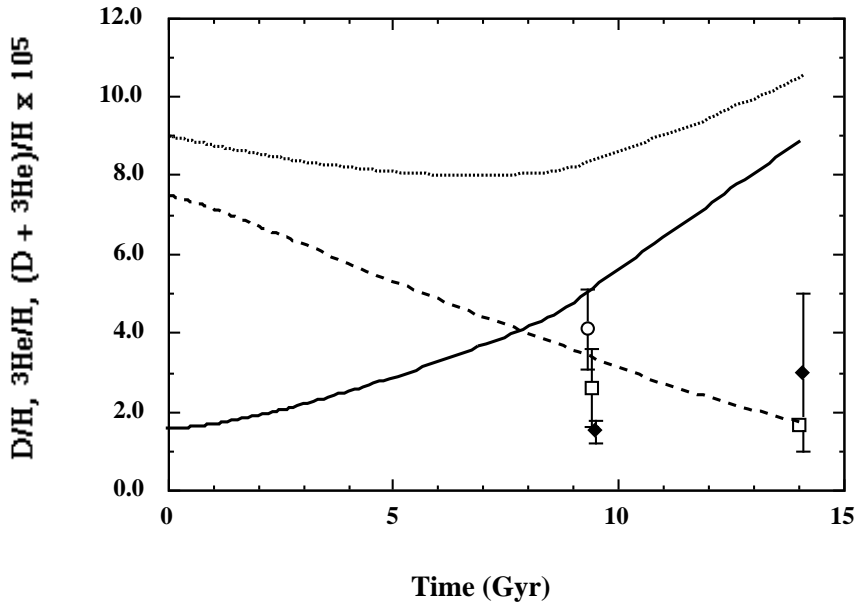}
\vspace{-7truecm}
\baselineskip=2ex
\vskip 3in
\caption { The evolution of D/H and 
\he3/H and (D+\he3)/H with time in units of $10^{-5}$. The assumed
primordial abundance of D/H is 7.5 $\times 10^{-5}$. The solid curve
shows the evolution of \he3/H, the dashed curve for D/H and the dotted
curve for the sum (D+\he3)/H. The diamonds show the data for
\he3, the open squares for deuterium and the open circle for the sum.}
\label{evol0}
\end{figure}

\begin{figure}
\hspace{0.5truecm}
\epsfysize=14truecm
\epsfbox{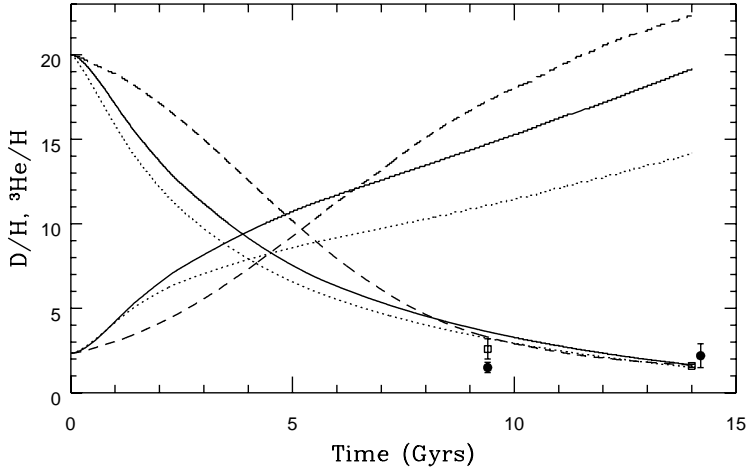}
\vspace{-7truecm}
\baselineskip=2ex
\caption { As in Figure \protect\ref{evol0}. The primordial abundance of
D/H in this case was chosen to be 2 $\times 10^{-4}$. }
\label{evol1}
\end{figure}

Generally, implementation of the \he3 yield in Eq. (\ref{it}) in chemical
evolution models leads to an overproduction of \he3/H particularly at the
solar epoch \cite{orstv,galli}.  For example, in Figure \ref{evol0},
the evolution of D and \he3 is shown for a model in which only a
modest amount of deuterium is destroyed.  Namely, by a factor of 5,
from D/H = 7.5 $\times 10^{-5}$ to a present value of $\sim 1.5 \times
10^{-5}$.  However, due to the production of \he3 in low mass stars, \he3
is greatly overproduced relative to the solar value. This problem is
compounded in models with an intense period of D destruction. In Scully
et al. \cite{scov}, a dynamically generated supernovae wind model was
coupled to models of galactic chemical evolution with the aim of reducing
a primordial D/H abundance of 2 $\times 10^{-4}$ to the present ISM value
without overproducing heavy elements and  remaining consistent with the
other observational constraints typically  imposed on such models.
In Figure \ref{evol1}, the evolution of D/H and 
\he3/H is
shown as a function of time in several representative models
with significant deuterium destruction factors (see ref \cite{scov} for
details).  However,
as one can plainly see, \he3 is grossly overproduced (the deuterium data is
represented by squares and \he3 by circles). 

\begin{figure}
\hspace{0.5truecm}
\epsfysize=14truecm\epsfbox{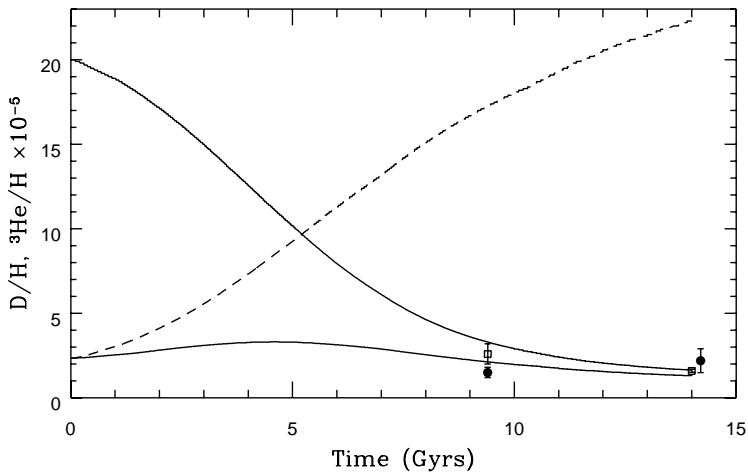}
\vspace{-7truecm}
\baselineskip=2ex
\caption { The evolution of D/H and \he3/H with time using the reduced
\he3 yields of ref. \protect\cite{bm} The dashed curve is the same as in
Figure
\protect\ref{evol1}, using standard \he3 yields.}
\label{evol2}
\end{figure}

The overproduction of \he3 relative to the solar meteoritic value seems to be a
generic feature of chemical evolution models when \he3 production in low mass
stars is included. This result appears to be independent of the 
chemical evolution
model and is directly related to the assumed stellar yields of \he3.
It has recently been suggested that at least some low mass
stars may indeed be net destroyers of \he3 if one includes
the effects of extra mixing below the conventional convection zone
in low mass stars on the red giant branch \cite{char,bm}. The extra  
mixing does not take place for stars which do not undergo a helium core
flash (i.e. stars $>$ 1.7 - 2 M$_\odot$ ).  Thus stars with masses {\it less than}
1.7 M$_\odot$ are responsible for the \he3 destruction. 
Using the yields of Boothroyd and Malaney \cite{bm}, it was shown \cite{osst}
that these reduced \he3 yields in low mass stars can account for the
relatively low solar and present day \he3/H abundances observed.
In fact, in some cases, \he3 was underproduced.  To account for the \he3 evolution
and the fact that some low mass stars must be producers 
of \he3 as indicated by the
planetary nebulae data, it was suggested that the new yields apply
only to a fraction (albeit large) of low mass stars \cite{osst,gal}. 
The corresponding evolution \cite{osst} of 
D/H and \he3/H is shown in Figure \ref{evol2}.

The models of chemical evolution discussed above indicate that it is
possible to destroy significant amounts of deuterium and remain
consistent with chemical evolutionary constraints.  To do so however, comes
with a price. Large deuterium destruction factors require substantial
amounts of stellar processing, which at the same time produce heavy
elements.  To keep the heavy element abundances in the Galaxy in check,
significant Galactic winds enriched in heavy elements must be
incorporated.  In fact there is some evidence that enriched winds were
operative in the early Galaxy.  In the X-ray cluster satellites observed
by Mushotzky et al. \cite{mush1} and  Loewenstein and Mushotzky
\cite{mush2} the mean oxygen abundance was found to be roughly half solar.
This corresponds to a near solar abundance of heavy elements in the
inter-Galactic medium, where apparently little or no star formation has
taken place. 

\begin{figure}
\hspace{0.5truecm}
\epsfysize=14truecm\epsfbox{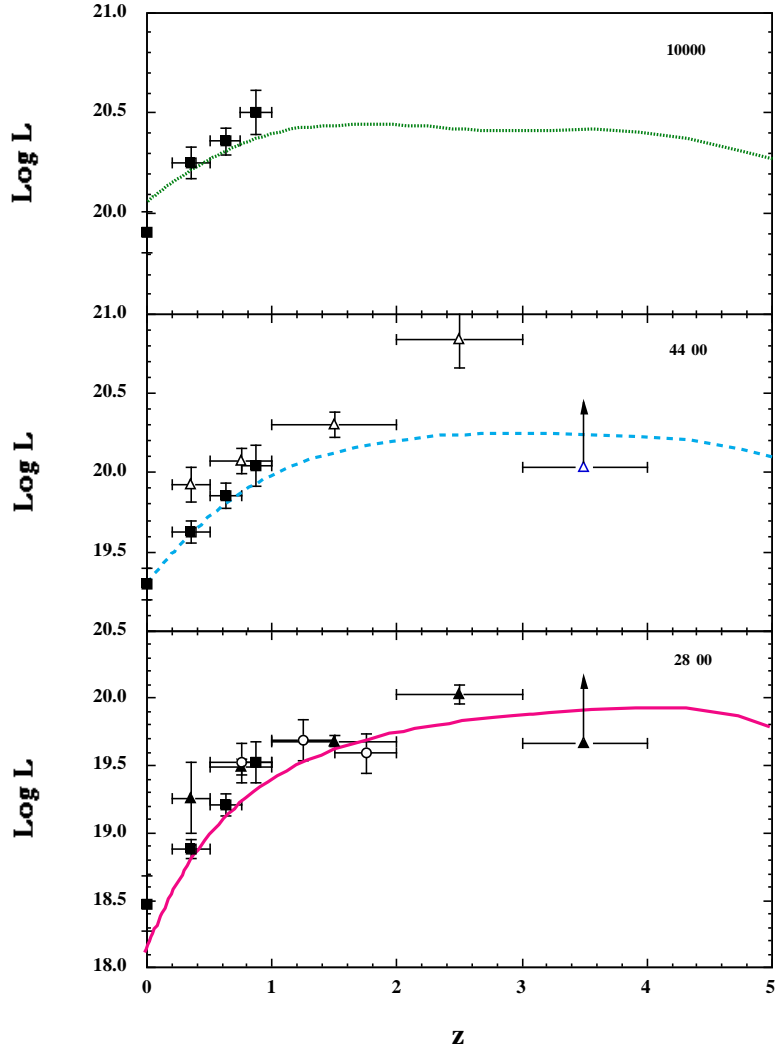}
%\vspace{-7truecm}
\baselineskip=2ex
\caption { The tricolor luminosity  densities (UV, B and IR) at
$\lambda = 0.28, 0.44$ and 1.0 $\mu$m, in units of (h/.5)
WHz$^{-1}$Mpc$^{-3}$ as a function of redshift for a model shown in
\protect\ref{evol2} which destroys significant amounts of D/H. The data
are taken from
\protect\cite{cce}.}
\label{fig:cova}
\end{figure}

If our Galaxy is typical in the Universe, then the models of the type
discussed above would indicate that the luminosity density of the
Universe at high redshift should also be substantial augmented relative
to the present. Recent observations of the luminosity density at high
redshift \cite{cce}  are making it possible for the first time
to test models of cosmic chemical evolution. 
The high redshift observations, are very discriminatory with
respect to a given SFR \cite{cova}.  Models in which the star formation
rate is proportional to the gas mass fraction (these are common place in
Galactic chemical evolution) have difficulties to fit the multi-color
data from $z = 0$ to 1.  This includes many of the successful Galactic
infall models. In contrast, 
models with  a steeply decreasing SFR  are
favored.  In Figure \ref{fig:cova}, the predicted luminosity
density based on the model with evolution shown in Figure \ref{evol2} from
\cite{scov}, as compared with the observations (see ref.
\cite{cova} for details).

While it would be premature to conclude that all models with large
deuterium destruction factors are favored, it does seem that models which
do fit the high redshift data destroy significant amounts of D/H.  On the
other hand, we can not exclude models which destroy only a small amount
of D/H as Galactic models of chemical evolution. In this case, however the
evolution of our Galaxy is anomalous with respect to the cosmic average.
If the low D/H measurements \cite{quas2,bty}  hold up, then it would
seem that our Galaxy also has an anomalously high  D/H abundance.  That
is we would predict in this case that the present cosmic abundance of
D/H is significantly lower than the observed ISM value.  If the high D/H
observations  \cite{quas1,quas3,webb} hold up, we could conclude that our
Galaxy is indeed representative of the cosmic star formation history.

\section{Likelihood Analyses}

Monte Carlo techniques have proven to be a useful form of analysis for big
bang nucleosynthesis \cite{kr,kk,hata1}. An
analysis of this sort was performed \cite{fo}
using only \he4 and \li7. Two
elements are sufficient for not only constraining the one parameter 
($\eta$) theory of BBN, but also for testing for consistency. 
The procedure begins by establishing likelihood functions for the theory and
observations. For example, for \he4, the theoretical likelihood 
function takes the
form
\beq
L_{\rm BBN}(Y,Y_{\rm BBN}) 
  = e^{-\left(Y-Y_{\rm BBN}\left(\eta\right)\right)^2/2\sigma_1^2}
\label{gau}
\eeq
where $Y_{\rm BBN}(\eta)$ is the central value for the \he4 mass fraction
produced in the big bang as predicted by the theory at a given value of $\eta$.
$\sigma_1$ is the uncertainty in that  value derived from the Monte Carlo
calculations \cite{hata1} and is a measure of the theoretical 
uncertainty in the
big bang calculation. Similarly one can write down an expression for the
observational likelihood function. Assuming Gaussian errors,
the likelihood function for the observations would
take a form similar to that in (\ref{gau}).

\begin{figure}
\hspace{0.5truecm}
\epsfysize=7truecm\epsfbox{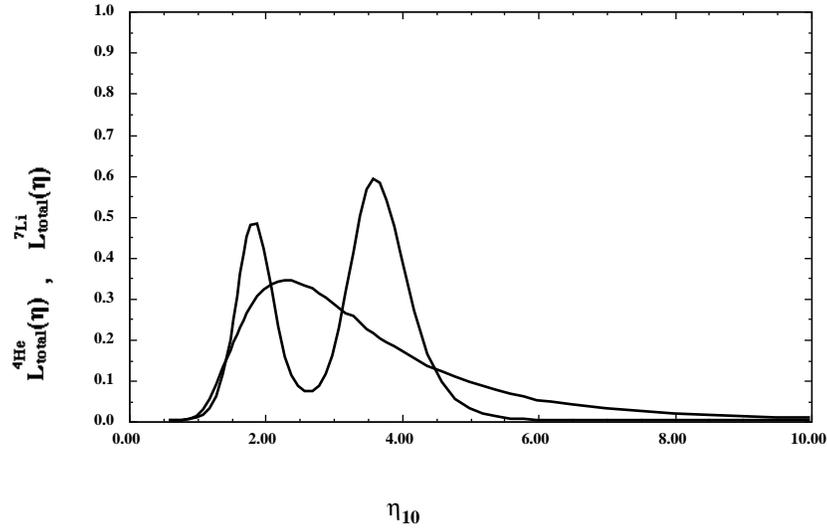}
%\vspace{-11.0truecm}
\caption { Likelihood distribution for each of \he4 and
\li7, shown as a  function of $\eta$.  The one-peak structure of the \he4
curve corresponds to the monotonic increase of $Y_p$ with $\eta$, while
the two peaks for \li7 arise from the minimum in the \li7 abundance prediction.}
\label{fig:fig1}
\end{figure}

A total likelihood 
function for each value of $\eta$ is derived by
convolving the theoretical
and observational distributions, which for \he4 is given by
\beq
{L^{^4{\rm He}}}_{\rm total}(\eta) = 
\int dY L_{\rm BBN}\left(Y,Y_{\rm BBN}\left(\eta\right)\right) 
L_{\rm O}(Y,Y_{\rm O})
\label{conv}
\eeq
An analogous calculation is performed \cite{fo} for \li7. 
The resulting likelihood
functions from the observed abundances given in Eqs. (\ref{he4}) 
  and (\ref{li})
is shown in Figure \ref{fig:fig1}. As one can see 
there is very good agreement between \he4 and \li7 in the range
of $\eta_{10} \simeq$ 1.5 -- 5.0. The double peaked nature of the \li7
likelihood function is due to the presence of a minimum in the
 predicted lithium abundance.  For a given observed value of \li7, there
are two likely values of $\eta$.

\begin{figure}
\hspace{0.5truecm}
\vskip .25in
\epsfysize=7truecm\epsfbox{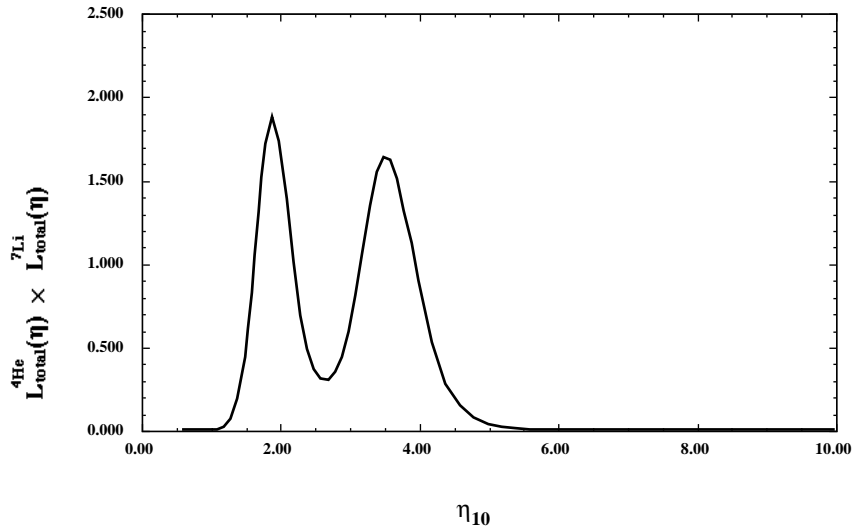}
%\vspace{-11.0truecm}
\baselineskip=2ex
\caption { Combined likelihood for simultaneously fitting \he4 and \li7,
as a function of $\eta$.}
\label{fig:fig2}
\end{figure}

The combined likelihood, for fitting both elements simultaneously,
is given by the product of the two functions in Figure \ref{fig:fig1}
and is shown in Figure \ref{fig:fig2}. The 95\% CL region covers the
range $1.55 < \eta_{10} < 4.45$, with the two peaks occurring at
$\eta_{10} = 1.9$ and 3.5. This range corresponds to values 
$\Omega_B$ between
\beq
0.006 < \Omega_B h^2 < .016
\label{omega}
\eeq

\begin{figure}
\hspace{0.5truecm}
\epsfysize=7truecm\epsfbox{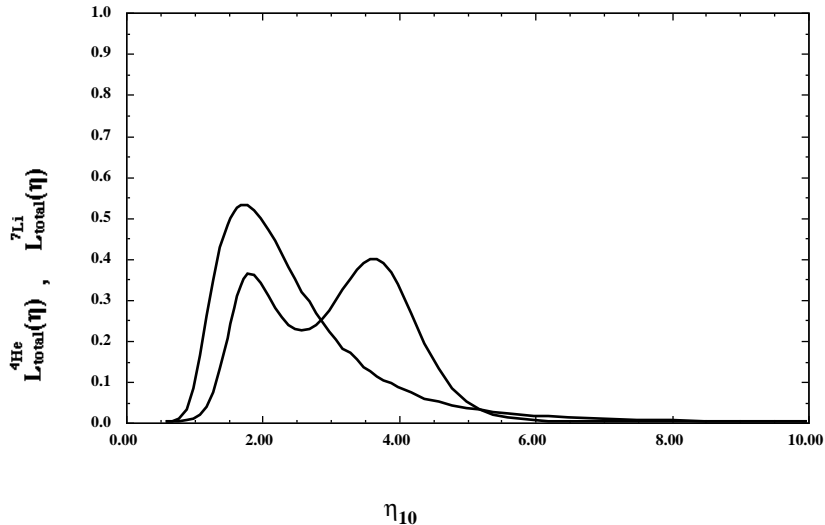}
%\vspace{-11.0truecm}
\caption { As in Figure \protect\ref{fig:fig1} with a lower value of
$Y_p = 0.234 \pm 0.002 \pm 0.005$.}
\label{old1}
\end{figure}

\begin{figure}
\hspace{0.5truecm}
\vskip .15in
\epsfysize=7truecm\epsfbox{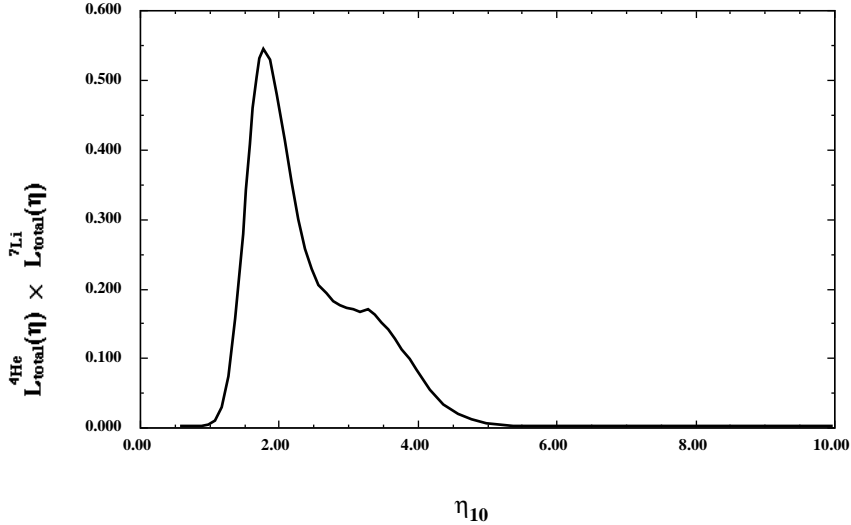}
%\vspace{-11.0truecm}
\baselineskip=2ex
\caption { Combined likelihood for simultaneously fitting \he4 and \li7,
as a function of $\eta$ from from Figure \protect\ref{old1}.}
\label{old2}
\end{figure}

For the lower value of $Y_p = 0.234 \pm 0.002 \pm 0.005$ as quoted in
\cite{ost3}, the \he4 peak is shifted to slightly lower values of $\eta$
and sits on top of the low-$\eta$ peak as shown in Figure \ref{old1}. 
(The difference in the \li7 likelihood distribution is due to the assumed
uncertainty in the \li7 abundance which is slightly higher than that in
Figure \ref{fig:fig1}.) The combined likelihood in this case is shown in
Figure \ref{old2}. From Figure \ref{old2} it is clear that \he4 overlaps
the lower (in $\eta$) \li7 peak, and so one expects that 
there will be concordance
in an allowed range of $\eta$ given by the overlap region.  
This is what one finds in Figure \ref{old2}, which does
show concordance and gives a preferred value for $\eta$, 
$\eta_{10}  = 1.8^{+2.4}_{-.4}$ corresponding to (at 95\% CL)
$\Omega_B h^2 = .007^{+.008}_{-.002}$.

Thus,  we can conclude that 
the abundances of 
\he4 and \li7 are consistent, and select an $\eta_{10}$ range which
overlaps with (at the 95\% CL) the longstanding favorite
 range around $\eta_{10} = 3$.
Furthermore, by finding concordance  
using only \he4 and \li7, we deduce that
if there is problem with BBN, it must arise from 
D and \he3 and is thus tied to chemical evolution or the stellar evolution of
\he3. The most model-independent conclusion is that standard
BBN  with $N_\nu = 3$ is not in jeopardy.

\begin{figure}
\hspace{0.5truecm}
\epsfysize=7truecm\epsfbox{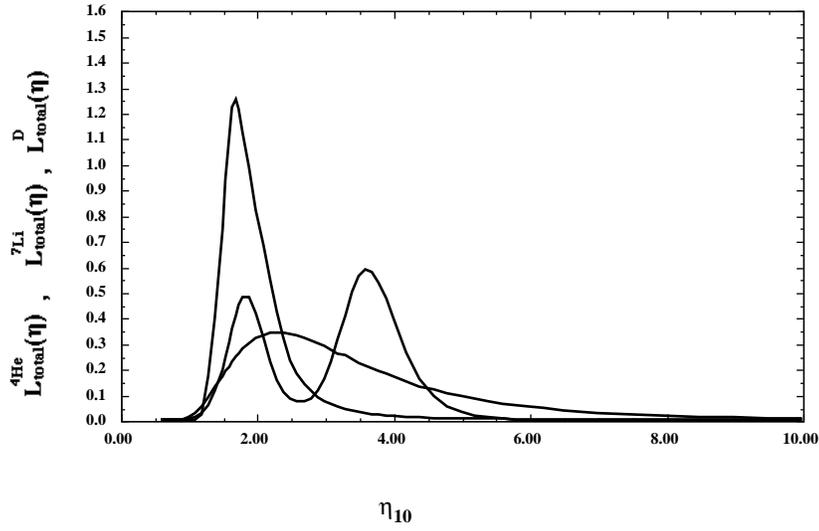}
%\vspace{-11.0truecm}
\baselineskip=2ex
\caption{ As in Figure \protect\ref{fig:fig1}, with the addition of the
likelihood  distribution for D/H assuming ``high" D/H.}
\label{D1}
\end{figure}

\begin{figure}
\hspace{0.5truecm}
\vskip .35in
\epsfysize=7truecm\epsfbox{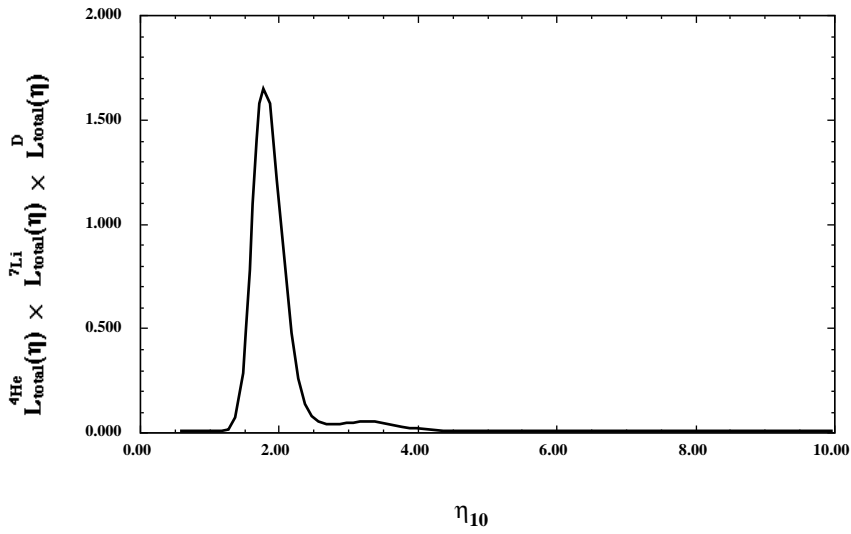}
%\vspace{-11.0truecm}
\baselineskip=2ex
\caption { Combined likelihood for simultaneously fitting 
\he4 and \li7, and D 
as a function of $\eta$ from Figure \protect\ref{D1}.}
\label{D2}
\end{figure}

 It is interesting to
compare the results from the likelihood function of \he4 and \li7 with
that of D/H.  
Since  D and \he3 are monotonic functions of $\eta$, a prediction for 
$\eta$, based on \he4 and \li7, can be turned into a prediction for
D and \he3.  
 The corresponding 95\% CL ranges are D/H  $= (4.3 - 25)  \times
10^{-5}$ and \he3/H $= (1.2 - 2.6)  \times 10^{-5}$.
If we did have full confidence in the measured value of D/H in 
quasar absorption
systems, then we could perform the same statistical analysis 
using \he4, \li7, and
D. To include D/H, one would
proceed in much the same way as with the other two light elements.  We
compute likelihood functions for the BBN predictions as in
Eq. (\ref{gau}) and the likelihood function for the observations.
These are then convolved as in Eq.  (\ref{conv}).  Using
D/H = $(2.0 \pm 0.5) \times 10^{-4}$ as indicated in the high
D/H systems, we can plot the three likelihood functions including
$L^{{\rm D}}_{\rm total}(\eta)$  in Figure \ref{D1}.
  It is indeed startling how the three peaks, for
D, \he4 and \li7 are in excellent agreement with each other.  In Figure
\ref{D2},  the combined distribution is shown.
We now  have a very clean distribution and prediction 
for $\eta$, $\eta_{10}  = 1.8^{+1.6}_{-.3}$ corresponding to $\Omega_B h^2
= .007^{+.005}_{-.001}$.  
The absence of any overlap with the high-$\eta$ peak of the \li7
distribution has considerably lowered the upper limit to $\eta$. 
Overall, the concordance limits in this case are dominated by the 
deuterium likelihood function.

% --------------------------------------
% Figure 4
% --------------------------------------

%\baselineskip=3ex

% --------------------------------------
% Figure 5
% --------------------------------------

\begin{figure}
\hspace{0.5truecm}
\epsfysize=7truecm\epsfbox{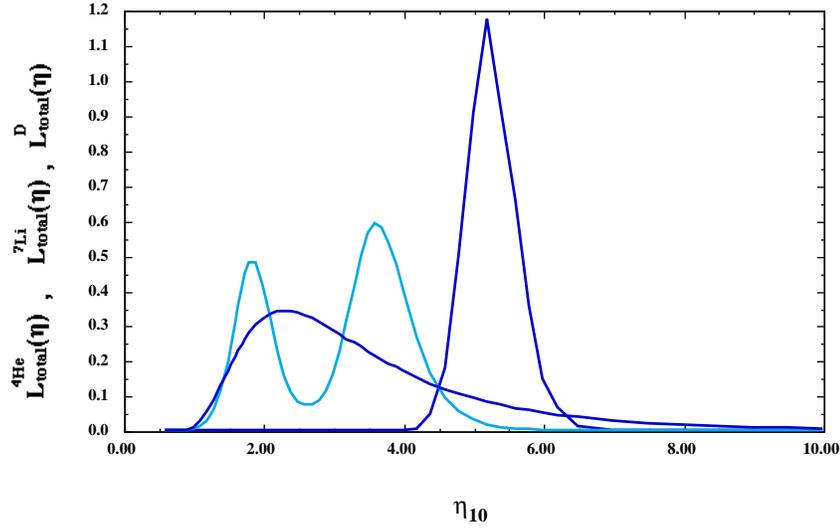}
%\vspace{-11.0truecm}
\baselineskip=2ex
\caption { As in Figure \protect\ref{D1}, with the likelihood 
distribution for low D/H. }
\label{Dt1}
\end{figure}

\begin{figure}
\hspace{0.5truecm}
\vskip .05in
\epsfysize=7truecm\epsfbox{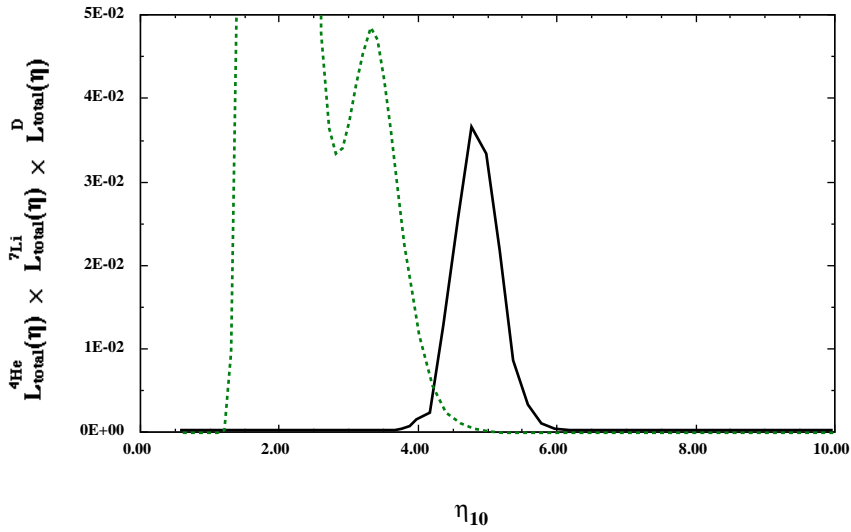}
%\vspace{-11.0truecm}
\baselineskip=2ex
\caption { Combined likelihood for simultaneously fitting 
\he4 and \li7, and  low D/H
as a function of $\eta$. The dashed curve represents the combined distribution
shown in Figure \protect\ref{D2}.}
\label{Dt2}
\end{figure}

 If instead, we assume that the low value \cite{bty}
of D/H = $(3.4 \pm 0.3) \times 10^{-5}$ is the primordial abundance,
then we can again compare the likelihood distributions as in Figure
\ref{D1}, now substituting the low D/H value. As one can see from Figure
\ref{Dt1}, there  is now hardly any overlap between the D and the \li7
and \he4 distributions.  The combined distribution shown in Figure \ref{Dt2} is
compared with that in Figure \ref{D2}. Though one can not use this likelihood
analysis to prove the correctness of the high
D/H measurements or the incorrectness of the low D/H measurements,
the analysis clearly shows the difference in compatibility between the
two values of D/H and the observational determinations of \he4 and \li7.
To {\em make} the low D/H measurement compatible, one would have to argue
for a shift upwards in \he4 to a primordial value of 0.247 (a shift by 0.009)
which is not warranted
at this time by the data, and a \li7 depletion factor of 
about 2, which is close to recent upper limits to the amount of depletion
\cite{cv,pin}.

% --------------------------------------
% Figure 6
% --------------------------------------

%\baselineskip=3ex

% --------------------------------------
% Figure 7
% --------------------------------------

The implications of the resulting predictions from big bang nucleosynthesis
on dark matter are clear.  First, if $\Omega = 1$ (as predicted by 
inflation), and $\Omega_B \la 0.1$ which is certainly a robust
conclusion based on D/H, then non-baryonic dark matter is a necessity.
Second, on the scale of small groups of galaxies, 
which are expected to sample the dark matter in galactic halos, $\Omega \ga
0.05$. This value can be compared with the best estimate for $\Omega_B$
from equation (\ref{omega}). The low $\eta$ peak in Figure
\ref{fig:fig2} corresponds to $\Omega_B h^2 = 0.007$ which for $h = 1/2$
gives $\Omega_B = 0.028$.  In this event,
some non-baryonic dark matter in galactic halos is required. On the other
hand, the high $\eta$ peak in \ref{fig:fig2} corresponds to $\Omega_B
h^2 = 0.013$ which for $h = 1/2$ gives $\Omega_B = 0.051$ and may be
consistent with halo densities (but only just). If we include the
data from the high D/H measurements in QSO absorbers this conclusion is
unchanged, only the low $\eta$ peak survives, and some non-baryonic dark
matter is needed in galactic halos. In contrast
\cite{2d}, the low D/H measurements would imply that
$\Omega_B h^2 = 0.019$ allowing for the possibility that $\Omega_B
\simeq 0.08$.  In this case, no non-baryonic dark matter is
required in galactic halos.  However, I remind the reader that
the low D/H value of 3.4 $\times 10^{-5}$ is at present only barely consistent
with either the observations of
\he4 or \li7 and their interpretations as being primordial abundances.

\section{Constraints from BBN}

Limits on particle physics beyond the standard model are mostly sensitive to
the bounds imposed on the \he4 abundance. 
As is well known, the $^4$He abundance
is predominantly determined by the neutron-to-proton ratio just prior to
nucleosynthesis and is easily estimated assuming that all neutrons are
incorporated into \he4 (see Eq. (\ref{ynp})).
As discussed earlier, the neutron-to-proton
ratio is fixed by its equilibrium value at the freeze-out of 
the weak interaction rates at a temperature $T_f \sim 1$ MeV modulo the
occasional free neutron decay.  Furthermore, freeze-out is determined by the
competition between the weak interaction rates and the expansion rate of the
Universe
\begin{equation}
{G_F}^2 {T_f}^5 \sim \Gamma_{\rm wk}(T_f) = H(T_f) \sim \sqrt{G_N N} {T_f}^2
\label{comp} \label{freeze}
\end{equation}
where $N$ counts the total (equivalent) number of relativistic particle
species. The presence
of additional neutrino flavors (or any other relativistic species) at 
the time of nucleosynthesis increases the overall energy density
of the Universe and hence the expansion rate leading to a larger 
value of $T_f$, $(n/p)$, and ultimately $Y_p$.  Because of the
form of Eq. (\ref{comp}) it is clear that just as one can place limits
\cite{ssg} on $N$, any changes in the weak or gravitational coupling constants
can be similarly constrained (for a recent discussion see ref. \cite{co}).
In concluding this lecture, I will discuss the current constraint on
$N_\nu$ the number of particle species (in neutrino units) and the
limit on the strength of new interactions, if 3 right-handed (nearly massless)
neutrinos are assumed to exist.

In the past, \he3 (together with D) has stood out 
in its importance for BBN, because 
it  provided a (relatively large) lower limit for the baryon-to-photon
ratio \cite{ytsso}, $\eta_{10} > 2.8$. This limit for a long 
time was seen to be
essential because it provided the only means for bounding $\eta$ from below
and in effect allows one to set an upper limit on the number of neutrino
flavors \cite{ssg}, $N_\nu$, as well as other constraints on particle physics
properties. That is, the upper bound to $N_\nu$ 
is strongly dependent on the lower bound to
$\eta$.  This is easy to see: given an observed value for $Y_p$, for lower
$\eta$, the
\he4 abundance drops, allowing for a larger $N_\nu$, which would raise the \he4
abundance back to the observed value. However, for $\eta < 4 \times 10^{-11}$,
corresponding to
$\Omega_B h^2 \la .001-.002$, which is not too different from galactic mass
densities,  there is no
bound whatsoever on $N_\nu$ \cite{ossty}. 
Of course, with the improved data on
\li7, we do have lower bounds on $\eta$ which exceed $10^{-10}$.
In fact, despite the uncertainty in the D/H abundance in quasar
absorption systems, the high D/H values can certainly be regarded as an
upper limit to primordial D/H, which also yield a lower limit to $\eta$.

Because, of new observations of D and \he3, and the new theoretical
work on chemical evolution sparked by these observations, 
the bound on $N_\nu$ which is tied directly 
to these isotopes, should be called into question.
As described earlier, the limits due to \he3 are ultimately
tied to the assumed yields of low mass stars.
Using the reduced yields as depicted in Figure \ref{evol2}, consistent
values of $\eta < 2.8$ are certainly possible.    Ultimately, as I have
said repeatedly, D/H measurements in quasar absorption systems may soon
resolve this issue.  However, the lower values of $\eta$, 
relax the bounds on the number of neutrino flavors.

As discussed above, the limit on $N_\nu$ comes about via the 
change in the expansion rate given by the Hubble parameter,
\beq
H^2 = {8 \pi G \over 3} \rho = {8 \pi^3  G \over 90} [N_{\rm SM} 
+ {7 \over 8} \Delta N_\nu] T^4
\eeq
when compared to the weak interaction rates. Here $N_{\rm SM}$
refers to the standard model value for N. At $T \sim 1$ MeV,
$N_{\rm SM} = 43/4$. Additional degrees of freedom will 
lead to an increase in the freeze-out temperature eventually leading to
a higher \he4 abundance. In fact, one 
can parameterize the dependence of $Y$ on $N_\nu$ by 
\beq
Y = 0.2262 + 0.0131 (N_\nu - 3) + 0.0135 \ln \eta_{10} 
\label{YY}
\eeq
in the vicinity of $\eta_{10} \sim 2$.  Eq. (\ref{YY}) also shows
the weak (log) dependence on $\eta$. However, rather than use
(\ref{YY}) to obtain a limit, it is preferable to use 
the likelihood method.

\begin{figure}
\hspace{0.5truecm}
%\vspace{-10cm}
\epsfysize=15truecm\epsfbox{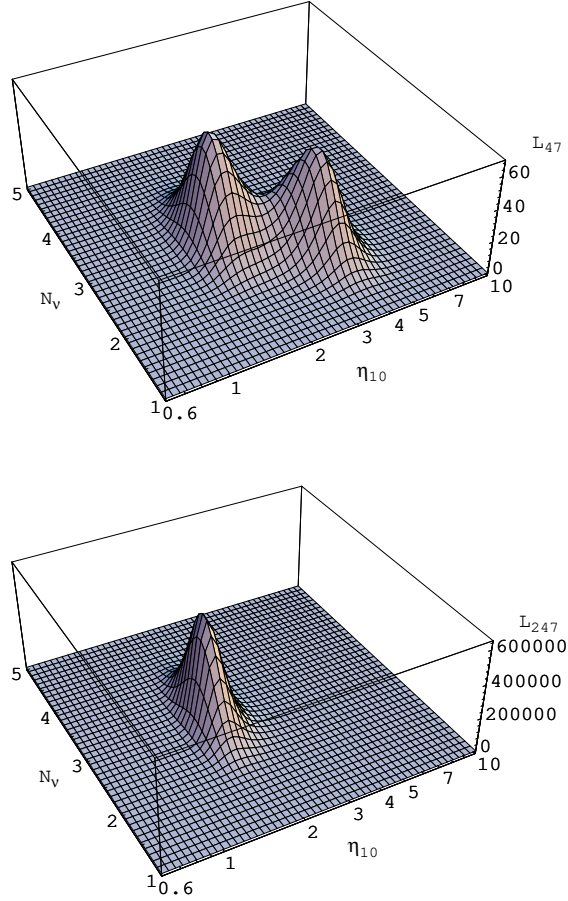}
%\vspace{-11.0truecm}
\baselineskip=2ex
\caption {The combined two-dimensional 
likelihood functions for simultaneously fitting 
\he4 and \li7 in the top panel, and including D in the lower one
as functions of both $\eta$ and $N_\nu$.}
\label{fig:fig1ai}
\end{figure}

 Just as \he4 and \li7 were sufficient to
determine a value for $\eta$,  a limit on $N_\nu$ can be obtained
as well \cite{fo,oth2}. The likelihood approach
utilized above can be extended to include $N_\nu$ as a free parameter.
Since the light element abundances can be computed as functions
of both $\eta$ and $N_\nu$,  the
likelihood function can be defined by \cite{oth2}
\beq
L_{\rm BBN}(Y,Y_{\rm BBN}) 
  = e^{-\left(Y-Y_{\rm BBN}\left(\eta,N_\nu\right)\right)^2/2\sigma_1^2}
\label{gau1}
\eeq
and \beq
{L^{^4{\rm He}}}_{\rm total}(\eta,N_\nu) = 
\int dY L_{\rm BBN}\left(Y,Y_{\rm BBN}\left(\eta,N_\nu\right)\right) 
L_{\rm O}(Y,Y_{\rm O})
\label{conv1}
\eeq
Again, similar expressions are needed for \li7 and D. 
A three-dimensional view of the combined likelihood functions \cite{oth2} 
is shown in Figure \ref{fig:fig1ai} which is based on the slightly
lower value of $Y_p$ as used in Figure \ref{old1}. For updated (but similar)
results see \cite{oth3}. In this case the high and low
$\eta$ maxima of Figure \ref{old2}, show up as peaks in the $L-\eta-N_\nu$
space ($L_{47}$ when D/H is neglected and $L_{247}$ when high D/H is
included). The peaks of the distribution as well as the
allowed ranges of $\eta$ and $N_\nu$ are  
%\newpage
%\vspace*{-7truecm}
\begin{figure}
\hspace{0.5truecm}
\epsfysize=15truecm\epsfbox{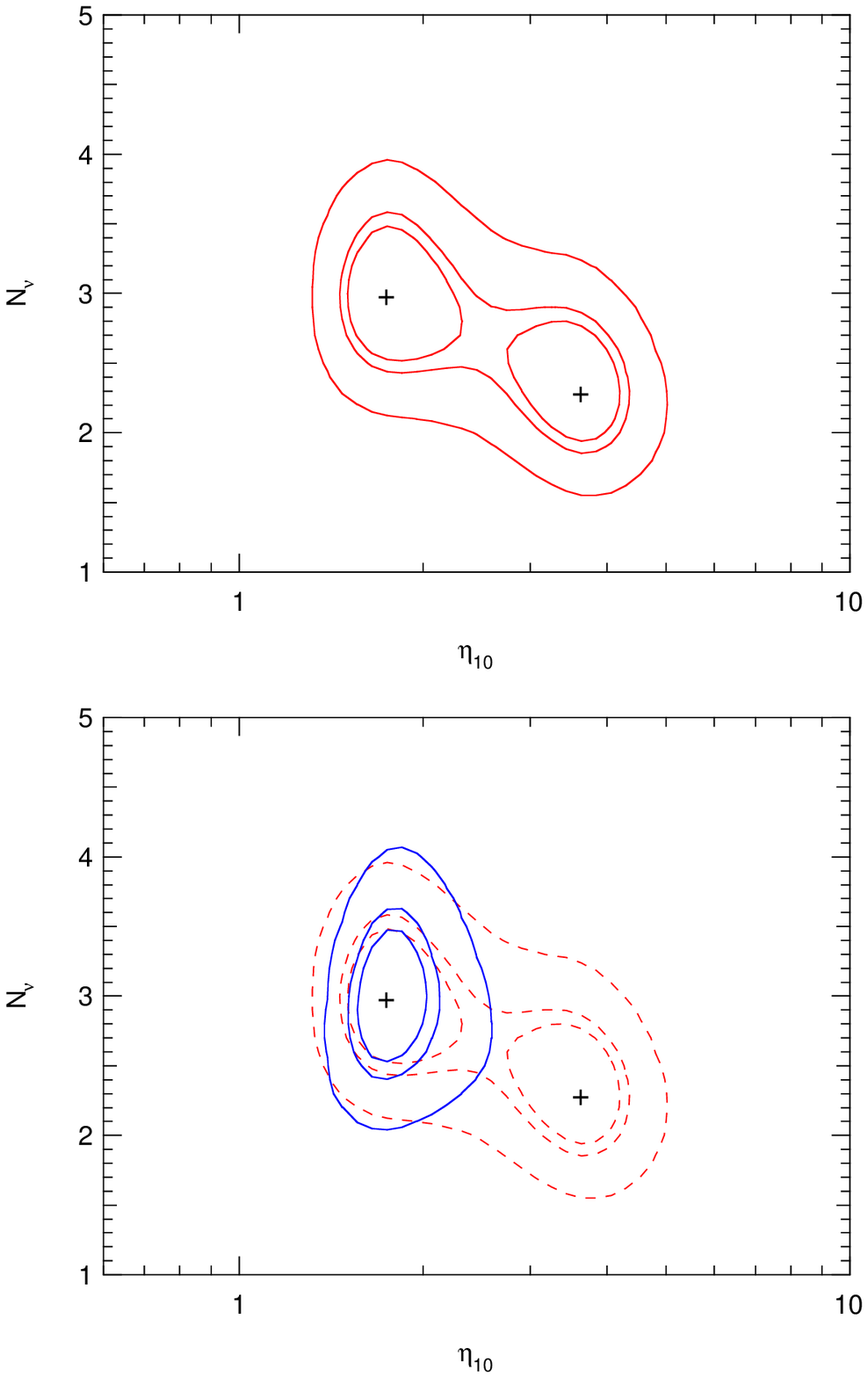}
%\vspace{-7.5truecm}
\caption{The top panel shows contours in the combined
		likelihood function for \he4 and \li7.
		The contours represent 50\% (innermost), 68\% and 95\% 
		(outermost) confidence levels.  The crosses mark the points 
		of maximum likelihood.  Also shown is the 
		equivalent result when D is included.}
\baselineskip=2ex
\label{fig:fig2ai}
\end{figure}
more easily discerned in the 
contour plot of Figure \ref{fig:fig2ai} which shows the 50\%,
68\% and 95\% confidence level contours in the two likelihood functions.  
The crosses show the location of the 
peaks of the likelihood functions.
$L_{47}$ peaks at $N_\nu=3.0$, $\eta_{10}=1.8$ (in agreement with 
our previous results \cite{fo}) and at $N_\nu=2.3$,
$\eta_{10}=3.6$.  The 95\% confidence level allows the following ranges
in $\eta$ and $N_\nu$
\beq
1.6\le N_\nu\le4.0  \qquad \qquad
%1.3\le~\eta_{10}~\le 5.0 
1.3\le\eta_{10}\le 5.0 
\eeq
Note however that the ranges in $\eta$ and $N_\nu$ are strongly
correlated as is evident in Figure \ref{fig:fig2ai}.
Since the deuterium likelihood function picks out a small range of values 
of $\eta$, largely independent of $N_\nu$, its effect on $L_{247}$ is 
to eliminate one of the two peaks in $L_{47}$. $L_{247}$ also 
peaks at $N_\nu=3.0$, $\eta_{10}=1.8$. 
In this case
the 95\% contour gives the ranges
\beq
2.0\le N_\nu\le4.1 \qquad \qquad
%1.4\le~\eta_{10}~\le 2.6 
1.4\le\eta_{10}\le 2.6 
\eeq
Finally, in Figure \ref{oth3fig}, the resulting 95 \% CL upper limit to 
$N_\nu$ is shown as a function of $Y_P$ for several different choices for
the primordial value of \li7/H \cite{oth3}.
\begin{figure}
\hspace{0.5truecm}
\epsfysize=7.5truecm\epsfbox{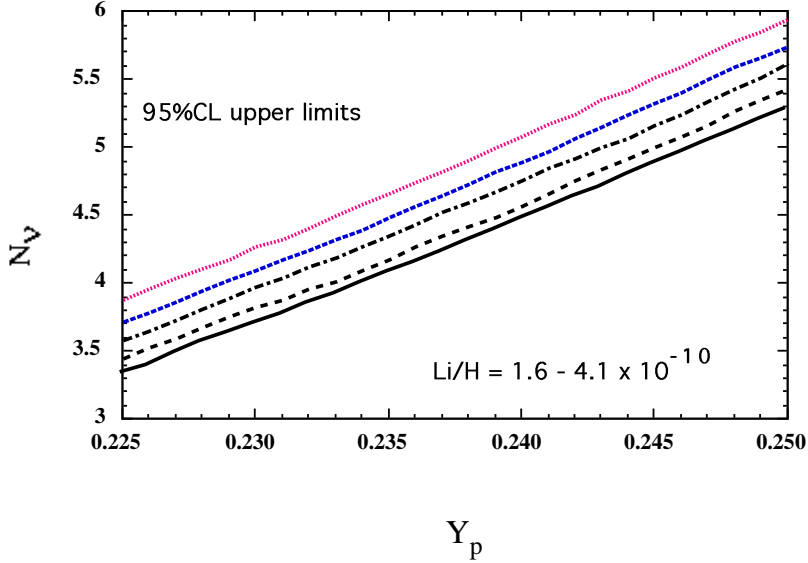}
%\vspace{-11.0truecm}
\caption { The 95 \% CL upper limit to $N_\nu$ as a function of $Y_p$
for (Li/H)$_p$ = 1.6, 2.0, 2.6, 3.2, and 4.1 $\times 10^{-10}$.}
\label{oth3fig}
\end{figure}

\begin{figure}
\hspace{0.5truecm}
\epsfysize=7truecm\epsfbox{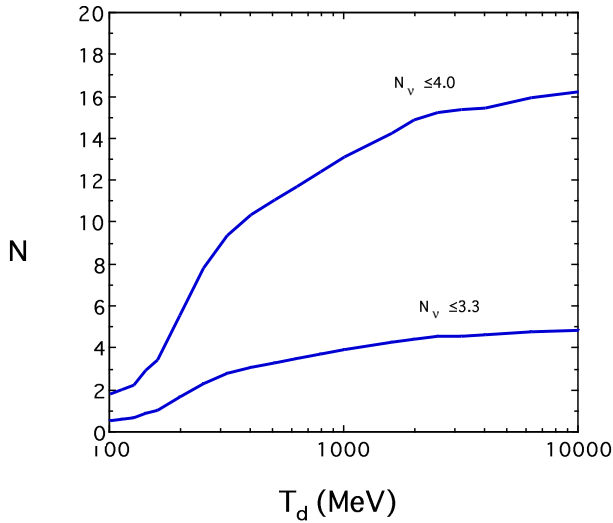}
\baselineskip=2ex
\caption { Limits on neutrino-like degrees of freedom.}
\label{last}
\end{figure}

One should recall that the limit derived above is not meant for neutrinos
in the strictest sense.  That is, the limit is only useful
when applied to additional particle degrees of freedom which 
necessarily do not couple to the Z$^o$. For very weakly interacting 
particles, one must take into account the reduced 
abundance of these particles at the time of nucleosynthesis\cite{oss}.  
As discussed in the first lecture, the number of neutrinos today is reduced
relative to the number of photons by
$(T_\nu/T_\gamma)^3  = 4/11$.  
For some new particle, $\chi$, which decoupled at $T_d > 1$ MeV, 
the same argument based on the conservation of entropy tells us that
\beq
({T_\chi \over T_\gamma})^3 = {43 \over 4 N(T_d)}
\label{decx}
\eeq
Thus we can translate the bound on $N_\nu$, which is really a bound 
on the additional energy density at nucleosynthesis
\beq
\Delta \rho = {\pi^2 \over 30} \left[ \sum g_B ({T_B \over T})^4
 + {7 \over 8} \sum g_F ({T_F \over T})^4 \right] T^4
\eeq
for additional boson states with $g_B$ degrees of freedom and
fermion states with $g_F$ degrees of freedom.
At nucleosynthesis $T = T_\nu = T_\gamma$ and the limit $N_\nu < 4$ 
becomes
\beq 
{8 \over 7} \sum {g_B \over 2} ({T_B \over T_\nu})^4
 +  \sum {g_F \over 2} ({T_F \over T})^4 < 1
\eeq
Such a limit would allow a single additional scalar
degree of freedom (which counts as ${4 \over 7}$) such as a majoron. 
On the other hand, 
in models with right-handed interactions, 
and three right-handed neutrinos, the
constraint is severe. 
The right-handed states must have decoupled early
enough to ensure $(T_{\nu_R}/T_{\nu_L})^4 < 1/3$. The
temperature of a decoupled state is easily determined from (\ref{decx}). 
 Three right-handed neutrinos would require  $N(T_d) \ga 25$, 
which from Figure \ref{mark} implies that $T_d
> 140$ MeV, conservatively assuming a QCD transition temperature
of 150 MeV. If right-handed interactions are mediated by additional gauge
interactions, associated with some scale $M_{Z'}$,
and if we assume that the right handed interactions scale
as $M_{Z'}^4$ with a standard model-like coupling, 
then the decoupling temperature of the right handed interactions
is related to $M_{Z'}$
by 
\beq
({{T_d}_R \over {T_d}_L})^3 \sim ({M_{Z'} \over M_Z})^4
\eeq
which for ${T_d}_L \sim 3$ MeV ( a more accurate value that the 
1 MeV estimate) and ${T_d}_L \ga 140$ MeV,
we find that the associated mass scale becomes $M_{Z'} \ga 1.6$ TeV!
In general this constraint is very sensitive to the BBN limit on $N_\nu$.
In Figure \ref{last}, the allowed number of neutrino degrees of freedom
are shown as a function of their decoupling temperature for the case
of $N_\nu < 4$ and $N_\nu < 3.3$, shown for comparison.

\section{Summary} 

To summarize on the subject of big bang nucleosynthesis, 
I would assert that one
can conclude that the present data on the abundances of the light element
isotopes are consistent with the standard model of big bang 
nucleosynthesis. Using
the the isotopes with the best data, \he4 and
\li7, it is possible to constrain the theory and obtain a best set of
values for the baryon-to-photon ratio of $\eta_{10}$ and the
corresponding  value for $\Omega_B h^2$ 
\beq
\begin{array}{ccccc}
1.55 & < & \eta_{10} & < &  4.45 \qquad 95\% {\rm CL} \nonumber \\
.006 & < & \Omega_B h^2 & < & .016  \qquad 95\% {\rm CL}
\label{res2}
\end{array}
\eeq
For $0.4 < h < 1$, we have a range $ .006 < \Omega_B < .10$.
This is a rather low value for the baryon density
 and would suggest that much of the galactic dark matter is
non-baryonic \cite{vc}. These predictions are in addition 
consistent with recent
observations of D/H in quasar absorption systems which show a high value.
Difficulty remains however, in matching the solar \he3 abundance, suggesting a
problem with our current understanding of galactic chemical evolution or the
stellar evolution of low mass stars as they pertain to \he3.

\end{document}